\documentclass[a4paper]{article}
\usepackage[utf8]{inputenc}
\usepackage[T1]{fontenc}
\usepackage[english]{babel}
\usepackage{amsmath,amsthm,amssymb,amsfonts}
\usepackage{xspace}
\usepackage{hyperref}
\usepackage{verbatim}
\usepackage{tikz}
\usepackage[procnumbered,linesnumbered,ruled,vlined]{algorithm2e}
\usepackage{tabularray}
\usepackage[hmargin=2.2cm,vmargin=3.8cm]{geometry}
\usepackage[textsize=footnotesize,color=green!40]{todonotes}
\usepackage{float}
\usepackage{stmaryrd}
\usepackage{thm-restate}
\usepackage{subcaption}
\usepackage{apxproof}
\usepackage[inline]{enumitem}
\usepackage{thm-restate}
\usepackage{hyperref}
\usepackage{cleveref}
\usepackage{authblk}

\hypersetup{
	colorlinks=true,
	citecolor=red,
	linkcolor=blue,
	filecolor=magenta,      
	urlcolor=black,
}
\newcommand{\shortversion}[1]{#1}
\newcommand{\longversion}[1]{}

\usetikzlibrary{arrows.meta}
\usetikzlibrary{calc}
\tikzstyle{noeud}=[circle,inner sep=2, minimum size =3 pt, line width = 1pt, draw=black, fill=white]
\newcommand{\nicearrow}{-{Latex[scale=1.1]}}
\newcommand{\convexpath}[2]{
	[   
	create hullnodes/.code={
		\global\edef\namelist{#1}
		\foreach [count=\counter] \nodename in \namelist {
			\global\edef\numberofnodes{\counter}
			\node at (\nodename) [draw=none,name=hullnode\counter] {};
		}
		\node at (hullnode\numberofnodes) [name=hullnode0,draw=none] {};
		\pgfmathtruncatemacro\lastnumber{\numberofnodes+1}
		\node at (hullnode1) [name=hullnode\lastnumber,draw=none] {};
	},
	create hullnodes
	]
	($(hullnode1)!#2!-90:(hullnode0)$)
	\foreach [
	evaluate=\currentnode as \previousnode using \currentnode-1,
	evaluate=\currentnode as \nextnode using \currentnode+1
	] \currentnode in {1,...,\numberofnodes} {
		-- ($(hullnode\currentnode)!#2!-90:(hullnode\previousnode)$)
		let \p1 = ($(hullnode\currentnode)!#2!-90:(hullnode\previousnode) - (hullnode\currentnode)$),
		\n1 = {atan2(\y1,\x1)},
		\p2 = ($(hullnode\currentnode)!#2!90:(hullnode\nextnode) - (hullnode\currentnode)$),
		\n2 = {atan2(\y2,\x2)},
		\n{delta} = {-Mod(\n1-\n2,360)}
		in 
		{arc [start angle=\n1, delta angle=\n{delta}, radius=#2]}
	}
	-- cycle
}

\usepackage[most]{tcolorbox}
\newcommand{\problemdef}[4]{
	\begin{tcolorbox}[width = \textwidth,colback=white,arc=0pt,outer arc=0pt,boxrule=0.7pt,left =0.5em,right=0em]\textsc{#1} #2		\\[2pt]
		\begin{tabular}{ @{}l p{0.84\textwidth} c }
			\textsf{Input:} & #3 \\[.5pt]
			\textsf{Problem:} & #4
		\end{tabular}
		\vspace{-0.25em}
	\end{tcolorbox}
}

\newcommand{\tempcc}{\textsc{Temporal Path Cover}\xspace}
\newcommand{\tempccshort}{\textsc{TPC}\xspace}

\newcommand{\tempdcc}{\textsc{Temporally Disjoint Path Cover}\xspace}
\newcommand{\tempdccshort}{\textsc{TD-PC}\xspace}
\newcommand{\kcover}{\textsc{Clique Cover}\xspace}

\newcommand{\ubp}{\textsc{Unary Bin Packing}\xspace}

\DeclareMathOperator{\tw}{tw}

\newcommand{\fprintrooted}{temporal rooted directed tree\xspace}
\newcommand{\fprintoriented}{temporal oriented tree\xspace}
\newcommand{\fprintdag}{temporal DAG\xspace}
\newcommand{\fprintrooteds}{temporal rooted directed trees\xspace}
\newcommand{\fprintorienteds}{temporal oriented trees\xspace}

\newcommand{\fprintdags}{temporal DAGs\xspace}
\newcommand{\connectivityGraph}{connectivity graph\xspace}

\DeclareMathOperator{\haut}{top}
\DeclareMathOperator{\bas}{bot}
\newcommand{\tmax}{t_{\max}}

\Crefname{claim}{Claim}{Claims}

\newcommand{\decisionpb}[4]{
	\begin{center}
		\noindent\framebox{\begin{minipage}{#4\textwidth}
				#1\\
				Instance: #2\\ 
				Question: #3
		\end{minipage}}%\vspace{0.5\baselineskip}
	\end{center}
}

\newcommand{\set}[1]{\left\{ #1\right\}}

\newtheorem{theorem}{Theorem}
\newtheorem{lemma}[theorem]{Lemma}
\newtheorem{definition}[theorem]{Definition}
\newtheorem{proposition}[theorem]{Proposition}
\newtheorem{observation}[theorem]{Observation}
\newtheorem{corollary}[theorem]{Corollary}
\newtheorem{claim}{Claim}%[theorem]

\newcommand{\ftodo}[2][]{\todo[color=blue!30,#1]{F: #2}} % Todos by Flo
\title{Algorithms and complexity for path covers of temporal DAGs: when is Dilworth dynamic?\thanks{This work was supported by the International Research Center "Innovation Transportation and Production Systems" of the I-SITE CAP 20-25 and by the ANR project GRALMECO (ANR-21-CE48-0004). Ralf Klasing's research was partially supported by the ANR project TEMPOGRAL (ANR-22-CE48-0001).}}
\author[1]{Dibyayan Chakraborty}
\author[2]{Antoine Dailly}
\author[2]{Florent Foucaud}
\author[3]{Ralf Klasing}
\affil[1]{School of Computing, University of Leeds, United Kingdom}
\affil[2]{Université Clermont-Auvergne, CNRS, Mines de Saint-Étienne, Clermont-Auvergne-INP, LIMOS, 63000 Clermont-Ferrand, France}
\affil[3]{Université de Bordeaux, Bordeaux INP, CNRS, LaBRI, UMR 5800, Talence, France}

\date{}

\begin{document}

\maketitle

\begin{abstract}
	
	A \emph{path cover} of a digraph is a collection of paths collectively containing the vertex set of the digraph. A path cover with minimum cardinality for a \emph{directed acyclic graph} can be found in polynomial time [Fulkerson, AMS '56; C{\'a}ceres et al., \textsc{SODA'22}].  Moreover, Dilworth's celebrated theorem on chain coverings of partially ordered sets equivalently states that the minimum size of a path cover of  a DAG is equal to the maximum size of a set of mutually unreachable vertices. In this paper,  we examine how far Dilworth's theorem can be extended to a ``\emph{dynamic}'' analogue of directed acyclic graphs.
	
	A temporal digraph has an arc set that changes over discrete time-steps. Furthermore, if the underlying digraph (\emph{i.e.},~the union of all the arc sets that appears at some point) is acyclic, then we have a temporal directed acyclic graph (or simply a temporal DAG). A temporal path is a directed path in the underlying digraph, such that the time-steps of arcs are strictly increasing along the path. Two temporal paths are temporally disjoint if they do not occupy any vertex at the same time. A \emph{temporal path cover} is a collection $\mathcal{C}$ of temporal paths that covers all vertices. Furthermore, $\mathcal{C}$ is a \emph{temporally disjoint path cover} if all temporal paths are pairwise temporally disjoint. In this paper, we study the computational complexities of the problems of finding a temporal (disjoint) path cover with minimum cardinality (denoted as \tempcc and \tempdcc). 
	
	We show that both \tempcc and \tempdcc are NP-hard even when the underlying DAG is planar, bipartite, subcubic, and there are only two arc-disjoint time-steps. Moreover, \tempdcc remains NP-hard even on temporal oriented trees. We also observe that natural temporal analogues of Dilworth's theorem on these classes of temporal DAGs do not hold.
	
	In contrast, we show that \tempcc is polynomial-time solvable on temporal oriented trees by a reduction to \kcover for (static undirected) \emph{weakly chordal} graphs (a subclass of perfect graphs for which \kcover admits an efficient algorithm). This highlights an interesting algorithmic difference between the two problems. Although it is NP-hard on temporal oriented trees, \tempdcc becomes polynomial-time solvable on temporal oriented lines and temporal \emph{rooted directed} trees. For all these positive algorithmic results, we also show that temporal analogues of Dilworth's theorem hold for the corresponding temporal graph classes.
	
	We also show that \tempcc and \tempdcc become efficiently solvable when the number of time-steps is bounded \emph{and} the underlying graph is close to a tree. More precisely, we show that \tempcc admits an XP time algorithm with respect to parameter $\tmax+tw$, where $\tmax$ is the maximum time-step, and $tw$ is the treewidth of the underlying static undirected graph. We also show that \tempdcc admits an FPT algorithm with respect to the same parameter.
\end{abstract}

%\newpage
%\setcounter{page}{1}
\section{Introduction}
\label{sec-intro}

A classic theorem of Dilworth from 1950~\cite{dilworth} states that in any partially ordered set (poset), the minimum number of chains required to cover all the elements is equal to the maximum size of an antichain. Dilworth's theorem is fundamental from the mathematical point of view; furthermore, an algorithmic proof (that enables to construct a chain cover and an antichain in polynomial time) was published by Fulkerson in 1956~\cite{fulkerson}. This theorem and its algorithmic form have many applications, not only in combinatorics, but also in various fields such as bioinformatics~\cite{PCmultiassembly}, scheduling~\cite{sched}, databases~\cite{database}, program testing~\cite{ntafos1979path}, etc.

A collection $\mathcal{P}$ of (resp.~pairwise vertex-disjoint) directed paths of a digraph $D$ is a \emph{path cover} (resp.~\emph{path partition}) of $D$ if all vertices of $D$ are contained in some path of $\mathcal{P}$. Dilworth's theorem can be restated in an equivalent form, equating the minimum cardinality of path covers on directed acyclic graphs (DAGs) and the maximum size of a set of pairwise ``unreachable'' vertices, or \emph{antichain} vertices~\cite{dagPC2,dagPC,FFbook}.

\begin{theorem}[Dilworth \cite{dilworth}]\label{thm:dil-path}
	For any DAG $D$, the minimum number of paths that cover its vertex set, is equal to the maximum size of an antichain of $D$.
\end{theorem}

Fulkerson~\cite{fulkerson} showed that finding a minimum-size path cover of a DAG can be done in polynomial time. Moreover, it is known that finding a minimum-size path partition can also be done in polynomial time for arbitrary DAGs~\cite[Probl.\,26-2]{CLRS3}. Improving the best known algorithms for path cover and partitions of DAGs still form an active field of research, see for example~\cite{dagPC2,dagPC,CC14,dagPC2019} for some recent results.

The notions of directed paths and path covers naturally extends to \emph{temporal (di)graphs}. Informally, the arc set of a temporal digraph changes over discrete time-steps and \emph{labels} of an arc are the time-steps where the arc appears. Temporal (di)graphs have been extensively studied in the two last decades, with contributions from and applications to various fields, see~\cite{CFQS12survey,H15-colloquium, KK15,temporal-transitive,M15intro,SH19book}. A \emph{temporal path} of a digraph is a path that traverses edges appearing at strictly increasing time-steps. The asymmetric nature of temporal paths has motivated many recent algorithmic works on related reachability or path problems on temporal graphs, such as~\cite{temporal-fastreach,AMS19,temporal-paths,CHMZ21,KKK02-TPP,temporal-eulerian}. 

Two temporal paths are \emph{temporally disjoint} if they do not occupy a vertex at the same time-step. This definition was introduced by Klobas et al.~\cite{temporalDP1} and has since then garnered attention in the graph algorithmic community~\cite{temporalDP2}. Even though the above notion was introduced in the context of temporal undirected graphs, it naturally extends to temporal digraphs and motivates the corresponding covering problems. The objective of \tempcc (resp. \tempdcc) is to cover an input temporal digraph by a minimum number of temporal paths (resp.~temporally disjoint paths). 

\medskip\noindent\textbf{Main objectives.} In this paper, we initiate the algorithmic study of \tempcc and \tempdcc and focus on \emph{temporal directed acyclic graphs} (or simply, temporal DAGs). A temporal digraph is a \emph{temporal DAG} if the union of all arcs across all time-steps induces a (static) DAG. We say that a temporal digraph satisfies the \emph{Dilworth property} (resp. \emph{temporally disjoint Dilworth property}, or \emph{TD-Dilworth property} for short) if the largest size of a \emph{temporal antichain} (understood as a set of pairwise unreachable vertices) is equal to the smallest size of a temporal path cover (resp. temporally disjoint path cover). The main goals of this paper are the following:

\begin{enumerate}
	\item[(a)] Determine classes of temporal DAGs satisfying the (TD-)Dilworth property.
	
	\item[(b)] Study the computational complexities of \tempcc and \tempdcc on temporal digraphs. 
\end{enumerate}

\medskip \noindent\textbf{Practical motivations.} A first motivation is multi-agent-based decision-making (a well-studied problem from artificial intelligence~\cite{SSFK0WLA0KB19,security}) in a temporal setting, such as for coral reef protection~\cite{coralreef} or crime prevention in transportation networks~\cite{escape-transport}. In this setting, the temporal DAG can model a decision-making process, where the vertices represent the states of an environment. Agents navigate the DAG, an arc representing an agent's move from one state to another. As the situation is varying over time, a move may only be available at specific time-steps. A path in this DAG thus represents the overall activity of an agent. In this setting, \tempcc represents the situation where a set of $k$ agents need to cover all the possible states. In \tempdcc, the agents must also avoid each other, and cannot cover the same state at the same time, a scenario described as \emph{vertex-conflicts} in the literature~\cite{SSFK0WLA0KB19}.

Another natural application is \emph{multi-robot path planning}~\cite{robots1,robots2}. Imagine the setting where $k$ robots are assigned the task of exploring a hazardous facility. Since the facility changes over time, it is modeled as a temporal digraph. If the facility digraph does not contain directed cycles, it is modeled by a temporal DAG (for example, if the facility is inherently directed from a start area towards a target area). The exploration path of a robot can be modeled by a temporal path. Now, \tempcc corresponds to the situation where the robots need to explore the whole facility, while for \tempdcc, the robots also cannot be simultaneously at the same location.

\medskip

%\subsection{Our results}
\noindent\textbf{Our results.} We begin by formally defining the problems studied in this paper. 
\problemdef{\tempcc}{(\tempccshort)}{A temporal digraph $ D$, an integer $k$.}{Does there exist a set $\mathcal{C}$ of $k$ temporal paths in $ D$ such that every vertex of $ D$ is covered by some path of $\mathcal{C}$?}

\problemdef{\tempdcc}{(\tempdccshort)}{A temporal digraph $ D$, an integer $k$.}{Does there exist a set $\mathcal{C}$ of $k$ temporally disjoint temporal paths in $ D$ such that every vertex of $ D$ is covered by some path of $\mathcal{C}$?}

We observe that in general, temporal DAGs do not have the Dilworth property (see \Cref{fig-introDAGNotDilworth}). Then, we prove the following negative result. 

\begin{restatable}{theorem}{dag}
	\label{thm-dag}
	\tempcc and \tempdcc are NP-hard on \fprintdags, even if the input is planar, bipartite, subcubic, of girth 10, uses only one time label per arc, and every label is either $1$ or $2$.
\end{restatable}

A temporal directed acyclic graph $D$ is a \emph{temporal oriented tree} if the underlying directed graph of $D$ is a tree. On the positive side, we prove the following. 

\begin{restatable}{theorem}{oriented}
	\label{thm-orientedTrees} There is an $\mathcal{O}(\ell n^2 + n^3)$-time algorithm for	\tempcc on \fprintorienteds with $n$ vertices and at most $\ell$ many labels per arc. Furthermore, \fprintorienteds satisfy the Dilworth property.
\end{restatable}

We briefly describe the technique we use for proving \Cref{thm-orientedTrees}.
Two vertices of a temporal digraph are \emph{temporally connected} if they are covered by the same temporal path. The \emph{connectivity graph} of a temporal digraph $D$ is an undirected (static) graph whose vertex set is the same as that of $D$, and whose edge set consists of all pairs of temporally connected vertices. To prove the above theorem, we show that the connectivity graph of a \fprintoriented is a \emph{weakly chordal graph}~\cite{Hayward85} (a subclass of \emph{perfect graphs}). We show \tempcc can be reduced to \textsc{Clique Cover} on weakly chordal graphs. The above observation, combined with the Weak Perfect Graph Theorem (proved by Lov{\'a}sz~\cite{lovasz1972normal}), proves that \fprintorienteds satisfy the Dilworth property. Moreover, the existing  $\mathcal{O}(nm)$-time algorithm~\cite{HaywardSS07} to compute a minimum clique cover of a weakly chordal graph (having $n$ vertices and $m$ edges) completes the proof of \Cref{thm-orientedTrees}. 
Our proof gives interesting structural information on the interaction between temporal paths in temporal oriented trees. Interestingly, another important class of perfect graphs plays an important role in connection with Dilworth's theorem and its translation to the setting of static DAGs: the class of comparability graphs, see~\cite[Chapter 5.7]{GolumbicBook}. In our case, there does not appear to be any connection to comparability graphs.

On the other hand, \fprintorienteds do not satisfy the TD-Dilworth property (see \Cref{fig-introTreeNotDilworth} for an example). Then, we prove the following negative result. 
\begin{restatable}{theorem}{orientnp}
	\label{thm-orientedTreesNP}
	\tempdcc is NP-hard on \fprintorienteds.
\end{restatable}

\begin{figure}
	\centering
	\captionsetup[subfigure]{justification=centering}
	\begin{subfigure}[b]{0.45\linewidth}
		\centering
		\scalebox{0.8}{
			\begin{tikzpicture}
				\node[noeud,fill=black] (0) at (0,0) {};
				\node[noeud] (1) at (2,0) {};
				\node[noeud] (2) at (4,0) {};
				\node[noeud] (3) at (6,0) {};
				
				\draw[line width = 0.25mm] \convexpath{0,1}{0.24cm};
				\draw[line width = 0.25mm] \convexpath{2,3}{0.24cm};
				
				\draw[\nicearrow] (0)to node[midway,below,yshift=-2mm]{3} (1);
				\draw[\nicearrow] (1)to node[midway,above,yshift=-1mm]{2} (2);
				\draw[\nicearrow] (2)to node[near start,above,yshift=2mm]{1} (3);
				
				\draw[\nicearrow,bend left] (0)to node[midway,above]{1} (3);
				\draw[\nicearrow,bend right] (1)to node[midway,below]{1} (3);
				\draw[\nicearrow,bend left] (0)to node[midway,below,yshift=1mm]{2} (2);
			\end{tikzpicture}
		}
		\caption{A \fprintdag not having the Dilworth property.}
		\label{fig-introDAGNotDilworth}
	\end{subfigure}
	\hfil
	\begin{subfigure}[b]{0.45\linewidth}
		\centering
		\scalebox{0.8}{
			\begin{tikzpicture}
				\node[noeud] (s1) at (0,3) {};
				\node[noeud] (s2) at (1,3) {};
				\node[noeud] (sk) at (3,3) {};
				\node[noeud] (u) at (1.5,1.5) {};
				\node[noeud] (t1) at (0,0) {};
				\node[noeud] (t2) at (1,0) {};
				\node[noeud] (tk) at (3,0) {};
				
				\draw[line width = 0.25mm] \convexpath{s1,u,t1,u,s1}{0.24cm};
				\fill[white] \convexpath{s1,u,t1,u,s1}{0.23cm};
				\draw[line width = 0.25mm] (s2) circle (0.24cm);
				\draw[line width = 0.25mm] (sk) circle (0.24cm);
				\draw[line width = 0.25mm] (t2) circle (0.24cm);
				\draw[line width = 0.25mm] (tk) circle (0.24cm);
				
				\node[noeud,fill=black] (s1) at (0,3) {};
				\node[noeud,fill=black] (s2) at (1,3) {};
				\node[noeud,fill=black] (sk) at (3,3) {};
				\node[noeud] (u) at (1.5,1.5) {};
				\node[noeud] (t1) at (0,0) {};
				\node[noeud] (t2) at (1,0) {};
				\node[noeud] (tk) at (3,0) {};
				
				\draw (2,3) node {$\mathbf{\ldots}$};
				\draw (2,0) node {$\mathbf{\ldots}$};
				
				\draw[\nicearrow] (s1)to node[midway,left,xshift=-2mm,yshift=-2mm]{1} (u);
				\draw[\nicearrow] (s2)to node[midway,right,yshift=1mm]{1} (u);
				\draw[\nicearrow] (sk)to node[midway,right,yshift=-1mm]{1} (u);
				\draw[\nicearrow] (u)to node[midway,left,xshift=-2mm,yshift=2mm]{2} (t1);
				\draw[\nicearrow] (u)to node[midway,right,yshift=-1mm]{2} (t2);
				\draw[\nicearrow] (u)to node[midway,right,yshift=1mm]{2} (tk);
				
				\draw (s1) node[above,yshift=2mm] {$s_1$};
				\draw (s2) node[above,yshift=2mm] {$s_2$};
				\draw (sk) node[above,yshift=2mm] {$s_k$};
				\draw (t1) node[below,yshift=-2mm] {$t_1$};
				\draw (t2) node[below,yshift=-2mm] {$t_2$};
				\draw (tk) node[below,yshift=-2mm] {$t_k$};
				\draw (u) node[right,xshift=2mm] {$c$};
			\end{tikzpicture}
		}
		\caption{A \fprintoriented not having the TD-Dilworth property.}
		\label{fig-introTreeNotDilworth}
	\end{subfigure}
	\caption{A minimum-size (temporally disjoint) temporal path cover is shown, vertices in a maximum-size temporal antichain are in black.}
\end{figure}

To find classes that satisfy the TD-Dilworth property, we study {\em temporal oriented lines} (that is, where the underlying digraph is an oriented path) and \emph{\fprintrooteds}. A tree is a \emph{rooted directed tree} if it is an oriented tree with a single source vertex called the \emph{root}. We prove the following result

\begin{restatable}{theorem}{rooted}
	\label{thm-rootedTrees}
	\tempcc and \tempdcc can be solved in time:
	\vspace*{-2mm}
	\begin{enumerate}
		\item[(a)] $\mathcal{O}(\ell n)$ on temporal oriented lines;
		\vspace*{-2mm}
		\item[(b)] $\mathcal{O}(\ell n^2)$ on \fprintrooteds;
	\end{enumerate}
	\vspace*{-2mm}
	where $\ell$ is the maximum number of labels per arc and $n$ is the number of vertices. Furthermore, both classes satisfy the TD-Dilworth property.
\end{restatable}

Note that some related problems remain NP-hard for temporal lines, such as \textsc{Temporally Disjoint Walks}~\cite{klobas2023interference}. \Cref{thm-rootedTrees}(a) shows that this is not the case here. To prove \Cref{thm-rootedTrees}(b), we begin by constructing a temporal path cover before transforming it into a temporally disjoint one of the same size. This is in contrast with general \fprintorienteds, for which, by Theorem~\ref{thm-orientedTreesNP}, such an approach is not possible.

As \tempdcc is NP-hard even on temporal oriented trees and on temporal DAGs with two time-steps, a natural question is what happens when the number of time-steps is small \emph{and} the underlying digraph is a tree. Motivated by this question, we study the case where both the number of time-steps and the treewidth of the underlying digraph are bounded (where we define the \emph{treewidth} of a temporal digraph as the treewidth of the underlying static undirected graph). We show that both problems become tractable in this setting. More precisely, we give a fixed-parameter tractable (FPT) algorithm for \tempdcc with treewidth and number of time-steps as parameters. The same technique gives an XP algorithm for \tempcc.

\begin{restatable}{theorem}{FPTTempDisjoint}
	\label{thm-fptTemporallyDisjoint}
	There is an algorithm for \tempdcc on general temporal digraphs that is FPT with respect to the treewidth of the underlying undirected graph and the maximum number of labels per arc. For \tempcc on general temporal digraphs, there is an XP algorithm for the same parameter.
\end{restatable}

See \Cref{tab-results} for a summary of our algorithmic results.

\shortversion{
	\begin{table}[t] \centering 
		\scalebox{1}{
			\begin{tblr}{
					colspec={Q[c,20em]Q[c,6em]Q[c,6em]},
					rows=m,
					vlines,
					hlines,
					hline{2}={2pt},
					vline{2}={2pt}
				}
				% \text{temporal graph class}  & \tempcc & \tempdcc & \tempcp \\ \hline\hline
				\text{temporal graph class}  & \tempccshort & \tempdccshort \\
				{\fprintdags (planar bipartite subcubic, girth~10, two arc-disjoint time-steps)} & \text{NP-c.}  & \text{NP-c.}  \\
				\fprintorienteds & \text{poly} & \text{NP-c.}  \\
				\fprintrooteds & \text{poly} & \text{poly} \\
				temporal oriented lines & \text{poly} & \text{poly}  \\
				{general temporal digraphs with \\ bounded treewidth and number of time-steps} & \text{poly (XP)} & \text{poly (FPT)} \\
		\end{tblr}}
		\caption{Summary of our algorithmic results. For all polynomial-time solvable classes of temporal DAGs, we also show that the Dilworth property (or TD-Dilworth property for \tempdccshort) holds.}
		\label{tab-results}
	\end{table}
}

\medskip

\noindent\textbf{Further related work.} Algorithms for solving several types of path and distance problems in temporal graphs have been developed, see for example~\cite{temporal-paths,KKK02-TPP,Wu-temporal-paths}. Recently, the problem \textsc{Temporally Disjoint Paths} was introduced in~\cite{temporalDP1}, as a generalization of the notorious \textsc{Disjoint Paths} problem (also known as \textsc{Linkage}). In \textsc{Temporally Disjoint Paths}, one is given a temporal graph with $k$ pairs of vertices called \emph{terminals}, and the goal is to find a set of $k$ pairwise temporally disjoint paths, each of them connecting one pair of terminals. \textsc{Temporally Disjoint Paths} is NP-hard, even for temporal lines and two paths~\cite{temporalDP1} or temporal stars~\cite{temporalDP2}, but becomes FPT for trees when parameterized by the number of paths~\cite{temporalDP1}. Algorithms that are FPT for certain structural parameters are given in~\cite{temporalDP2}.

\medskip

\noindent\textbf{Structure of the paper.} We start with the hardness result for \fprintdags (\Cref{thm-dag}) in \Cref{sec-dag}. We then prove our results for \fprintorienteds (\Cref{thm-orientedTrees} and \Cref{thm-orientedTreesNP}) in \Cref{sec-proof-Thm3,sec-oriented}. We prove \Cref{thm-rootedTrees}, the polynomial-time algorithms for special \fprintorienteds (\fprintrooteds and temporal oriented lines), in \Cref{sec-rooted}. We then prove our results for temporal digraphs of bounded treewidth and number of time-steps (\Cref{thm-fptTemporallyDisjoint}) in \Cref{sec-tw}. We conclude in \Cref{sec:conclu}.% Proofs of propositions, lemmas and claims marked with (*) are given in the Appendix.

\section{Preliminaries}

A \emph{temporal digraph} $\mathcal D=(V,A_1,\ldots,A_{\tmax})$ is given by a sequence of arc-sets representing $\tmax$ discrete \emph{time-steps} $\{1,\ldots,\tmax\}$, where an arc in $A_i$ is \emph{active} at time-step $i$~\cite{temporalDP1}. Let us denote by $D=(V,A)$, where $A=\cup_{i=1}^{\tmax} A_i$, the \emph{underlying digraph} of temporal digraph $\mathcal D=(V,A_1,\ldots,A_{\tmax})$ (sometimes called \emph{footprint (di)graph}~\cite{CFQS12survey}).
Equivalently, one can view the time-steps as an arc-labelling function $\lambda:A(D)\to 2^{[\tmax]}$, where $\lambda(\overrightarrow{xy})\subseteq [\tmax]$ is the set of time-steps where $\overrightarrow{xy}$ is active~\cite{KKK02-TPP}. In that case, we may denote the temporal digraph as $\mathcal D=(D,\lambda)$.
We say that a temporal digraph has a given property $\mathcal{P}$ (planarity, given girth...) if the undirected graph obtained by forgetting the orientation of the arcs of its underlying digraph has property $\mathcal{P}$.
For a given temporal digraph, we denote by $\ell$ the maximum number of labels per arc and by $n$ the number of vertices in the underlying digraph. 

For a (temporal) (di)graph $\mathcal{D}$ and subset $S$ of its vertices (resp. edges), $\mathcal{D}\setminus S$ denotes the (temporal) (di)graph obtained by removing the vertices (resp. edges) in $S$ from $\mathcal{D}$. 

In a temporal digraph, a \emph{temporal (directed) path} is a sequence $(v_1,v_2,t_1),(v_2,v_3,t_2),\ldots,\allowbreak (v_{k-1},v_k,t_{k-1})$ such that for any $i,j$ with $1\leq i<j\leq k$, $v_i\neq v_j$ and for any $i$ with $1\leq i\leq k-1$, $t_i<t_{i+1}$ and there is an arc $\overrightarrow{v_iv_{i+1}}$ at time-step $t_i$. These paths are sometimes called \emph{strict} in the literature.\footnote{For non-strict paths, the condition $t_i<t_{i+1}$ is replaced with $t_i\leq t_{i+1}$, but as argued in~\cite{temporalDP2}, the strict definition is more natural for applications where an agent cannot traverse an arbitrary number of arcs at once, this is why we chose this convention.} For a temporal path $P=(v_1,v_2,t_1),\ldots,\allowbreak (v_{k-1},v_k,t_{k-1})$, we denote by $V(P)$ the set $\cup_{i=1}^k \{v_i\}$ and by $A(P)$ the set $\cup_{i=1}^{k-1} \{ \overrightarrow{v_iv_{i+1}} \}$. 

The \emph{length} of a temporal path is the number of arcs it uses. 
We say that a temporal path $P=(v_1,v_2,t_1),\ldots,\allowbreak (v_{k-1},v_k,t_{k-1})$ \emph{occupies} vertex $v_i$ during the time interval $\{t_{i-1},\ldots,t_i\}$. Two temporal paths $P_1,P_2$ are \emph{temporally disjoint} if for all arcs $e_1\in A(P_1), e_2\in A(P_2)$ incident with a common end-vertex, the time-step of $e_1$ in $P_1$ is distinct from the time-step of $e_2$ in $P_2$.
In other words, two paths are temporally disjoint if they do not occupy the same vertex at the same time. A \emph{temporal path cover} (resp. \emph{temporally disjoint path cover}) of a temporal digraph $D$ is a collection of temporal paths (resp. temporally disjoint paths) that cover all vertices of $D$. Two vertices are \emph{temporally connected} in $D$ if there exists a temporal path between them. A \emph{temporal antichain} is a set of vertices that are pairwise not temporally connected. 

\begin{definition}
	A class $\mathcal{C}$ has the \emph{Dilworth property} (resp. \emph{TD-Dilworth property}) if the cardinality of the minimum temporal path cover (resp. temporally disjoint path cover) is equal to the maximum cardinality of a temporal anti-chain.
\end{definition}

A \emph{hole} of a static undirected graph is an induced cycle of length at least $5$, and an \emph{anti-hole} is the complement of a hole. A graph $G$ is \emph{weakly chordal} if it has no hole or anti-hole. A \emph{(minimum) clique cover} of a graph $G$ is a (minimum cardinality) set of complete subgraphs of $G$ that covers all vertices. A \emph{(maximum) independent set} of a graph $G$ is a (maximum cardinality) set of pairwise non-adjacent vertices. 
We shall use the following results for weakly chordal graphs.

\begin{theorem}[\cite{HaywardSS07,lovasz1972normal, SpinradS95}] \label{thm:weakly}
	Let $H$ be a weakly chordal graph with $n$ vertices and $m$ edges. Then, a minimum clique cover of $H$ can be found in $\mathcal{O}(nm)$-time. Furthermore, the maximum size of an independent set of $H$ equals the minimum size of a clique cover of $H$. 
\end{theorem}

\section{Temporal DAGs}
\label{sec-dag}

We provide a reduction (inspired by~\cite{MT07}) from a restricted variant of {\sc 3-Dimensional Matching} to prove the following.%\shortversion{ (proof deferred to the appendix due to space constraints)}.

\dag*

%\begin{toappendix}
%	\dag*
	%\begin{proof}[Proof of \Cref{thm-dag}]
	\begin{proof}
		We will reduce the {\sc Temporal (Disjoint) Path Cover} problem on \fprintdags from the {\sc 3-Dimensional Matching} problem.
		The reduction is inspired from~\cite{MT07}.%\rtodo{Florent, should we also argue what is different in our reduction? Actually, we did not have to add much...}\ftodo{no need to say anything!}\rtodo{I fully agree. Actually, I used the word {\em inspired} instead of {\em adapted}...}
		\decisionpb{{\sc 3-Dimensional Matching (3DM)}}{A set $S\subseteq X\times Y\times Z$, where $X$, $Y$, and $Z$ are disjoint sets having the same number $q$ of elements.}{Does $S$ contain a {\em perfect matching}, \emph{i.e.}, a subset $M\subseteq S$ such that $|M|=q$ and no two elements of $M$ agree in any coordinate?}{0.9}
		It is well-known that {\sc 3-Dimensional Matching} is NP-hard~\cite{GareyJ79}.
		
		Given an instance $I = (S, X \times Y \times Z)$ of {\sc 3DM},
		where $S=\{s_1\ldots,s_p\}$, $X=\{x_1,\ldots,x_q\}$, $Y=\{y_1,\ldots,y_q\}$ and $Z=\{z_1,\ldots,z_q\}$,
		we build an instance $\mathcal{D} = (V , A_1, A_2)$ of {\sc Temporal (Disjoint) Path Cover}, where $\mathcal{D}$ is a \fprintdag, as follows.
		
		To each triple $s_i =(x_{i,1},y_{i,2},z_{i,3}) \in S$, we associate a
		gadget $H(s_i)$ that consists of a collection
		$\{P^{i,1}, P^{i,2}, P^{i,3} \}$ of 3 directed vertex-disjoint paths of 3 vertices with
		$P^{i,r} =\{\overrightarrow{a_1^{i,r} a_2^{i,r}}, \overrightarrow{a_2^{i,r} a_3^{i,r}}\}$ for $r=1,2,3$; and the time labels are~1 for the arcs $\overrightarrow{a_1^{i,r} a_2^{i,r}}$ and~2 for the arcs $\overrightarrow{a_2^{i,r} a_3^{i,r}}$. We add to
		$H(s_i)$ the arcs $\overrightarrow{a_3^{i,1} a_3^{i,2}}$ and
		$\overrightarrow{a_3^{i,2} a_3^{i,3}}$,
		in order to form a 4th directed path of 3 vertices; and the time labels are~1 for the arcs $\overrightarrow{a_3^{i,1} a_3^{i,2}}$ and~2 for the arcs $\overrightarrow{a_3^{i,2} a_3^{i,3}}$. Finally, we add to $H(s_i)$ the arcs $\overrightarrow{a_2^{i,1} x_{i,1}}$, $\overrightarrow{a_2^{i,2} y_{i,2}}$ and  $\overrightarrow{a_2^{i,3} z_{i,3}}$, with the time label 2
		(see Figure~\ref{gadget-dags} for an illustration).
		
		\begin{figure}[h]
			\centering
			%	\resizebox{0.72\textwidth}{!}{%
				\begin{tikzpicture}[scale=.8]
					\node[noeud] (0) at (0,0) {$a_1^{i,1}$};
					\node[noeud] (1) at (2,0) {$a_2^{i,1}$};
					\node[noeud] (2) at (4,0) {$a_1^{i,2}$};
					\node[noeud] (3) at (6,0) {$a_2^{i,2}$};
					\node[noeud] (4) at (8,0) {$a_1^{i,3}$};
					\node[noeud] (5) at (10,0) {$a_2^{i,3}$};
					\node[noeud] (6) at (2,2) {$a_3^{i,1}$};
					\node[noeud] (7) at (6,2) {$a_3^{i,2}$};
					\node[noeud] (8) at (10,2) {$a_3^{i,3}$};
					\node[noeud] (9) at (2,-2) {$x_{i,1}$};
					\node[noeud] (10) at (6,-2) {$y_{i,2}$};
					\node[noeud] (11) at (10,-2) {$z_{i,3}$};
					
					%	\draw[line width = 0.25mm] \convexpath{0,1}{0.24cm};
					%	\draw[line width = 0.25mm] \convexpath{2,3}{0.24cm};
					%	\draw[line width = 0.25mm] \convexpath{4,5}{0.24cm};
					
					\draw[\nicearrow] (0)to node[midway,above]{1} (1);
					%	\draw[\nicearrow] (1)to node[midway,above,yshift=-1mm]{2} (2);
					%	\draw[\nicearrow] (2)to node[near start,above,yshift=2mm]{1} (3);
					\draw[\nicearrow] (2)to node[midway,above]{1} (3);
					%	\draw[\nicearrow] (4)to node[midway,below,yshift=-2mm]{1} (5);
					\draw[\nicearrow] (4)to node[midway,above]{1} (5);
					\draw[\nicearrow] (1)to node[midway,right]{2} (6);
					\draw[\nicearrow] (3)to node[midway,right]{2} (7);
					\draw[\nicearrow] (5)to node[midway,right]{2} (8);
					\draw[\nicearrow] (6)to node[midway,above]{1} (7);
					\draw[\nicearrow] (7)to node[midway,above]{2} (8);
					\draw[\nicearrow] (1)to node[midway,right]{2} (9);
					\draw[\nicearrow] (3)to node[midway,right]{2} (10);
					\draw[\nicearrow] (5)to node[midway,right]{2} (11);
					
					%	\draw[\nicearrow,bend left] (0)to node[midway,above]{1} (3);
					%	\draw[\nicearrow,bend right] (1)to node[midway,below]{1} (3);
					%	\draw[\nicearrow,bend left] (0)to node[midway,below,yshift=1mm]{2} (2);
				\end{tikzpicture}
				%}%
			\caption{The gadget $H(s_i)$.}
			\label{gadget-dags}
		\end{figure}
		
		The above construction yields a temporal digraph $\mathcal{D}$ on $9p + 3q$ vertices.
		Note that %all the temporal paths in $G$ are strict, 
		the  construction uses only 1 label per arc, and every label is either 1 or 2.
		
		We claim that there exists a perfect matching $M \subseteq S$ in $I$ if and only if there exists a temporal (disjoint) path cover (partition) of $\mathcal{D}$ of size $3p+q$.
		
		($\Rightarrow$) Let $M \subseteq S$ be a perfect matching in $S$, and consider the following collection of directed vertex-disjoint temporal length-2 paths in the gadget $H(s_i)$:
		$\{\overrightarrow{a_3^{i,1} a_3^{i,2}}, \overrightarrow{a_3^{i,2} a_3^{i,3}}\}$,
		$\{\overrightarrow{a_1^{i,1} a_2^{i,1}}, \overrightarrow{a_2^{i,1} x_{i,1}}\}$,  
		$\{\overrightarrow{a_1^{i,2} a_2^{i,2}}, \overrightarrow{a_2^{i,2} x_{i,2}}\}$,
		$\{\overrightarrow{a_1^{i,3} a_2^{i,3}}, \overrightarrow{a_2^{i,3} x_{i,3}}\}$
		if $s_i \in M$, 
		and $P^{i,1}, P^{i,2}, P^{i,3}$ if $s_i \not\in M$.
		As $M$ is a perfect matching in $S$, the collection of the temporal paths defined above constitute a vertex-disjoint (and thus temporally disjoint) temporal path cover of $\mathcal{D}$ of size $3p+q$. %\Cref{partition-gadget-1} and \Cref{partition-gadget-2}
		\Cref{partition-gadget-1,partition-gadget-2}
		illustrate the construction of the temporal path partition on
		$V(H(s_i))$ with respect to a given matching $M$ for {\sc 3DM}.
		
		\begin{figure}
			\centering
			\captionsetup[subfigure]{justification=centering}
			\begin{subfigure}[b]{0.48\textwidth}
				\centering
				\scalebox{0.58}{
					\begin{tikzpicture}
						\node[noeud] (0) at (0,0) {$a_1^{i,1}$};
						\node[noeud] (1) at (2,0) {$a_2^{i,1}$};
						\node[noeud] (2) at (4,0) {$a_1^{i,2}$};
						\node[noeud] (3) at (6,0) {$a_2^{i,2}$};
						\node[noeud] (4) at (8,0) {$a_1^{i,3}$};
						\node[noeud] (5) at (10,0) {$a_2^{i,3}$};
						\node[noeud] (6) at (2,2) {$a_3^{i,1}$};
						\node[noeud] (7) at (6,2) {$a_3^{i,2}$};
						\node[noeud] (8) at (10,2) {$a_3^{i,3}$};
						\node[noeud] (9) at (2,-2) {$x_{i,1}$};
						\node[noeud] (10) at (6,-2) {$y_{i,2}$};
						\node[noeud] (11) at (10,-2) {$z_{i,3}$};
						
						\draw[line width = 0.5mm] \convexpath{0,1,9,1,0}{0.6cm};
						\fill[white] \convexpath{0,1,9,1,0}{0.575cm};
						\draw[line width = 0.5mm] \convexpath{2,3,10,3,2}{0.6cm};
						\fill[white] \convexpath{2,3,10,3,2}{0.575cm};
						\draw[line width = 0.5mm] \convexpath{4,5,11,5,4}{0.6cm};
						\fill[white] \convexpath{4,5,11,5,4}{0.575cm};
						\draw[line width = 0.5mm] \convexpath{6,7,8}{0.6cm};
						
						\node[noeud] (0) at (0,0) {$a_1^{i,1}$};
						\node[noeud] (1) at (2,0) {$a_2^{i,1}$};
						\node[noeud] (2) at (4,0) {$a_1^{i,2}$};
						\node[noeud] (3) at (6,0) {$a_2^{i,2}$};
						\node[noeud] (4) at (8,0) {$a_1^{i,3}$};
						\node[noeud] (5) at (10,0) {$a_2^{i,3}$};
						\node[noeud] (6) at (2,2) {$a_3^{i,1}$};
						\node[noeud] (7) at (6,2) {$a_3^{i,2}$};
						\node[noeud] (8) at (10,2) {$a_3^{i,3}$};
						\node[noeud] (9) at (2,-2) {$x_{i,1}$};
						\node[noeud] (10) at (6,-2) {$y_{i,2}$};
						\node[noeud] (11) at (10,-2) {$z_{i,3}$};
						
						\draw[\nicearrow] (0)to node[midway,above]{1} (1);
						\draw[\nicearrow] (2)to node[midway,above]{1} (3);
						\draw[\nicearrow] (4)to node[midway,above]{1} (5);
						\draw[\nicearrow] (1)to node[midway,right]{2} (6);
						\draw[\nicearrow] (3)to node[midway,right]{2} (7);
						\draw[\nicearrow] (5)to node[midway,right]{2} (8);
						\draw[\nicearrow] (6)to node[midway,above]{1} (7);
						\draw[\nicearrow] (7)to node[midway,above]{2} (8);
						\draw[\nicearrow] (1)to node[midway,right]{2} (9);
						\draw[\nicearrow] (3)to node[midway,right]{2} (10);
						\draw[\nicearrow] (5)to node[midway,right]{2} (11);
					\end{tikzpicture}
				}
				\caption{$s_i \in M$}
				\label{partition-gadget-1}
			\end{subfigure}
			\hfill
			\begin{subfigure}[b]{0.48\textwidth}
				\centering
				\scalebox{0.58}{
					\begin{tikzpicture}
						\node[noeud] (0) at (0,0) {$a_1^{i,1}$};
						\node[noeud] (1) at (2,0) {$a_2^{i,1}$};
						\node[noeud] (2) at (4,0) {$a_1^{i,2}$};
						\node[noeud] (3) at (6,0) {$a_2^{i,2}$};
						\node[noeud] (4) at (8,0) {$a_1^{i,3}$};
						\node[noeud] (5) at (10,0) {$a_2^{i,3}$};
						\node[noeud] (6) at (2,2) {$a_3^{i,1}$};
						\node[noeud] (7) at (6,2) {$a_3^{i,2}$};
						\node[noeud] (8) at (10,2) {$a_3^{i,3}$};
						\node[noeud] (9) at (2,-2) {$x_{i,1}$};
						\node[noeud] (10) at (6,-2) {$y_{i,2}$};
						\node[noeud] (11) at (10,-2) {$z_{i,3}$};
						
						\draw[line width = 0.5mm] \convexpath{0,1,6,1,0}{0.6cm};
						\fill[white] \convexpath{0,1,6,1,0}{0.575cm};
						\draw[line width = 0.5mm] \convexpath{2,3,7,3,2}{0.6cm};
						\fill[white] \convexpath{2,3,7,3,2}{0.575cm};
						\draw[line width = 0.5mm] \convexpath{4,5,8,5,4}{0.6cm};
						\fill[white] \convexpath{4,5,8,5,4}{0.575cm};
						
						\node[noeud] (0) at (0,0) {$a_1^{i,1}$};
						\node[noeud] (1) at (2,0) {$a_2^{i,1}$};
						\node[noeud] (2) at (4,0) {$a_1^{i,2}$};
						\node[noeud] (3) at (6,0) {$a_2^{i,2}$};
						\node[noeud] (4) at (8,0) {$a_1^{i,3}$};
						\node[noeud] (5) at (10,0) {$a_2^{i,3}$};
						\node[noeud] (6) at (2,2) {$a_3^{i,1}$};
						\node[noeud] (7) at (6,2) {$a_3^{i,2}$};
						\node[noeud] (8) at (10,2) {$a_3^{i,3}$};
						\node[noeud] (9) at (2,-2) {$x_{i,1}$};
						\node[noeud] (10) at (6,-2) {$y_{i,2}$};
						\node[noeud] (11) at (10,-2) {$z_{i,3}$};
						
						\draw[\nicearrow] (0)to node[midway,above]{1} (1);
						\draw[\nicearrow] (2)to node[midway,above]{1} (3);
						\draw[\nicearrow] (4)to node[midway,above]{1} (5);
						\draw[\nicearrow] (1)to node[midway,right]{2} (6);
						\draw[\nicearrow] (3)to node[midway,right]{2} (7);
						\draw[\nicearrow] (5)to node[midway,right]{2} (8);
						\draw[\nicearrow] (6)to node[midway,above]{1} (7);
						\draw[\nicearrow] (7)to node[midway,above]{2} (8);
						\draw[\nicearrow] (1)to node[midway,right]{2} (9);
						\draw[\nicearrow] (3)to node[midway,right]{2} (10);
						\draw[\nicearrow] (5)to node[midway,right]{2} (11);
					\end{tikzpicture}
				}
				\caption{$s_i \not\in M$}
				\label{partition-gadget-2}
			\end{subfigure}
			\caption{Vertex partition of the gadget $H(s_i)$ into length-2 paths.}
			\label{partition-gadget}
		\end{figure}
		
		($\Leftarrow$) Assume that there exists a (temporally disjoint) path cover $\mathcal{C}$ of $\mathcal{D}$ of size $3p+q$. As $|V(\mathcal{D})| = 9q+3p$, $|\mathcal{C}|= 3p+q$ and every (temporal) path in $\mathcal{D}$ has length at most 2, all paths in $\mathcal{C}$ must have exactly length~2, and $\mathcal{C}$ is indeed a {\em partition} of $V(\mathcal{D})$ into length-2 paths. All length-2 paths in $\mathcal{D}$ are depicted in \Cref{partition-gadget-1,partition-gadget-2}. Hence, any path partition $\mathcal{C}$ of $\mathcal{D}$ must have, for each triple gadget, the path structure as depicted in either one of \Cref{partition-gadget-1,partition-gadget-2}, and there must be $q$ gadgets $H_1,\dots,H_q$  that are each covered by four vertex-disjoint temporal length-2 paths from $\mathcal{C}$, and $p-q$ gadgets $H_{q+1},\dots,H_p$ where the vertices $a_1^{i,r} a_2^{i,r}a_3^{i,r}$ (for $r=1,2,3$) are covered by three vertex-disjoint temporal length-2 paths from $\mathcal{C}$ and the vertices $x_{i,1},x_{i,2},x_{i,3}$ are not covered.
		Then, the triples $(x_{i,1},y_{i,2},z_{i,3})$ corresponding to $H_1,\ldots,H_q$ constitute a perfect matching in $S$.
		
		This completes the NP-hardness proof of {\sc Temporal (Disjoint) Path Cover} in \fprintdags. In order to show that {\sc Temporal (Disjoint) Path Cover} remains NP-hard in planar, bipartite, subcubic \fprintdags of girth 10, we apply the above proof, except that we start from a restriction of the three-dimensional matching problem, in which every element appears in either two or three triples, and the associated bipartite graph (formed by the elements and triples as its vertices, with edges connecting each element to the triples it belongs to) is planar and subcubic, denoted by {\sc Planar 3DM-3}. It is well-known that this restriction of {\sc 3DM} is still NP-hard~\cite{DyerF86}. Following a planar embedding of the bipartite graph associated to the instance of {\sc Planar 3DM-3}, one can obtain a planar enmbedding of the constructed graph.
		Note that the underlying DAG in the above reduction is bipartite, as it can be 2-colored as follows: vertices in $X$ and $Z$ are colored 1, vertices in $Y$ are colored 2, and then the coloring can be extended to the triple gadgets. Note as well that the shortest cycle in the underlying undirected graph has length~10.
	\end{proof}
%\end{toappendix}

We also show the following.%\shortversion{ (proof in appendix due to space constraints)}.

%\begin{restatable}{proposition}{transitiveTournament}
%	\label{prop-dilworthInDAGs}
%	There are \fprintdags (whose underlying digraph is a transitive tournament) that satisfy neither the Dilworth nor the TD-Dilworth property. Moreover, the ratio between the minimum-size temporal path cover and temporally disjoint path cover and the maximum-size temporal antichain can be arbitrarily large.%\rtodo{What about the \emph{TD-Dilworth property}?} A: Done! The optimal cover is vertex-disjoint.
%\end{restatable}

\begin{proposition}%[*]
	\label{prop-dilworthInDAGs}
	There are \fprintdags (whose underlying digraph is a transitive tournament) that satisfy neither the Dilworth nor the TD-Dilworth property. Moreover, the ratio between the minimum-size temporal path cover and temporally disjoint path cover and the maximum-size temporal antichain can be arbitrarily large.
\end{proposition}

%\begin{toappendix}
	%\transitiveTournament*
	
	%\begin{proof}[Proof of \Cref{prop-dilworthInDAGs}]
	\begin{proof}%[Proof of \Cref{prop-dilworthInDAGs}]
		Consider the temporal digraph $\mathcal{T}_n=(T_n,\lambda)$ %\rtodo{Why did you use \mathtt here and not \mathcal? it was a mistake} 
		where $T_n$ is the transitive tournament on vertices $u_1,\ldots,u_n$ and $\lambda(\overrightarrow{u_iu_j}) = n-(j+1)$ for all $i<j$. $\mathcal{T}_n$ is a \fprintdag, and since its underlying digraph is a transitive tournament, all the pairs of vertices are temporally connected, implying that the temporal antichain is of size~1. However, no temporal path can contain more than two vertices, and thus $\left\lceil \frac{n}{2} \right\rceil$ paths are needed to cover it. Hence, the gap between the maximum size of a temporal antichain and the minimum size of a temporal path cover can be as large as we want for a \fprintdag. Furthermore, the minimum-size \tempccshort is also vertex-disjoint (and thus temporally disjoint) if $n$ is even since the paths will never intersect; and if $n$ is odd, then at most two paths in a minimum-size \tempccshort can intersect in at most one vertex, thus we can reduce one of the two intersecting paths to cover only one vertex, giving us a \tempccshort of the same size with vertex-disjoint, and thus temporally disjoint, paths.
		This is depicted in \Cref{fig-fprintdag-transitive-tournament}. 
	\end{proof}
	
	\begin{figure}[h]
		\centering
		\captionsetup[subfigure]{justification=centering}
		\begin{subfigure}[b]{0.38\textwidth}
			\scalebox{0.8}{
				\begin{tikzpicture}
					\node[noeud] (0) at (0,0) {};
					\node[noeud] (1) at (2,0) {};
					\node[noeud] (2) at (4,0) {};
					\node[noeud] (3) at (6,0) {};
					
					\draw[line width = 0.25mm] \convexpath{0,1}{0.24cm};
					\draw[line width = 0.25mm] \convexpath{2,3}{0.24cm};
					
					\draw[\nicearrow] (0)to node[midway,below,yshift=-2mm]{3} (1);
					\draw[\nicearrow] (1)to node[midway,above,yshift=-1mm]{2} (2);
					\draw[\nicearrow] (2)to node[near start,above,yshift=2mm]{1} (3);
					
					\draw[\nicearrow,bend left] (0)to node[midway,above]{1} (3);
					\draw[\nicearrow,bend right] (1)to node[midway,below]{1} (3);
					\draw[\nicearrow,bend left] (0)to node[midway,below,yshift=1mm]{2} (2);
				\end{tikzpicture}
			}
			\caption{$n=4$}
		\end{subfigure}
		\unskip\ \vrule\ 
		\begin{subfigure}[b]{0.58\textwidth}
			\scalebox{0.8}{
				\begin{tikzpicture}
					\node[noeud] (0) at (0,0) {};
					\node[noeud] (1) at (2,0) {};
					\node[noeud] (2) at (4,0) {};
					\node[noeud] (3) at (6,0) {};
					\node[noeud] (4) at (8,0) {};
					
					\draw[line width = 0.25mm] \convexpath{0,1}{0.24cm};
					\draw[line width = 0.25mm] \convexpath{3,4}{0.24cm};
					\draw[line width = 0.25mm] (2) circle (0.24);
					
					\draw[\nicearrow] (0)to node[near end,above,yshift=2mm]{4} (1);
					\draw[\nicearrow,bend right] (0)to node[midway,above,yshift=-1mm]{3} (2);
					\draw[\nicearrow,bend left] (0)to node[midway,above]{2} (3);
					\draw[\nicearrow,bend right] (0)to node[midway,above,yshift=-1mm]{1} (4);
					
					\draw[\nicearrow] (1)to node[midway,above,yshift=-1mm]{3} (2);
					\draw[\nicearrow] (2)to node[midway,above,yshift=-1mm]{2} (3);
					\draw[\nicearrow] (3)to node[near start,above,yshift=2mm]{1} (4);
					
					\draw[\nicearrow,bend left] (1)to node[midway,above]{1} (4);
					\draw[\nicearrow,bend right] (2)to node[midway,above,yshift=-1mm]{1} (4);
					%\draw[\nicearrow,bend left] (1)to node[midway,below,yshift=1mm]{2} (3);
					\draw[\nicearrow,bend right] (1)to node[midway,above,yshift=-1mm]{2} (3);
				\end{tikzpicture}
			}
			\caption{$n=5$}
		\end{subfigure}
		\caption{$\mathcal{T}_n$, a \fprintdag with a maximum-size temporal antichain of size~1 and a minimum-size temporal path cover of size $\left\lceil \frac{n}{2} \right\rceil$, for $n=4,5$. }
		\label{fig-fprintdag-transitive-tournament}
	\end{figure}
%\end{toappendix}

\section{\tempcc on temporal oriented trees}
\label{sec-proof-Thm3}

In this section we prove the following theorem.

\oriented*

For the rest of this section, $\mathcal{T}=(T,\lambda)$ shall denote a temporal oriented tree with $n$ vertices and at most $\ell$-many labels per edge. We construct the \emph{\connectivityGraph of $\mathcal{T}$}, denoted by $G$, as follows: $V(G)=V(T)$ and $E(G) = \{ uv~|~u \neq v \mbox{ and } u \mbox{ and } v \mbox{ are temporally connected}\}$.
In other words, the \connectivityGraph of a \fprintoriented connects vertices that are temporally connected.
Observe that $G$ can be constructed in $\mathcal{O}(\ell n^2)$-time. %\rtodo{Should we add a brief (informal) explanation how this can be done in time $\mathcal{O}(\ell n^2)$? AD: We don't have much space, and it's quite obvious, so I'd say no.}
The next observation follows immediately from the definition.

\begin{observation}\label{obs:max-ind-antichain}
	A set $S$ of vertices of $\mathcal{T}$ is a temporal antichain if and only if $S$ induces an independent set in $G$.
\end{observation}

We have the following relationship between temporal paths in $\mathcal{T}$ and cliques in $G$.

\begin{lemma}%[*]
	\label{lem:min-clique-cover}
	Let $S$ be a set of vertices of $\mathcal{T}$. Then $S$ is contained in a temporal path in $\mathcal{T}$ if and only if $S$ is contained in a clique of $G$.
\end{lemma}

%\begin{toappendix}
	\begin{proof}%[Proof of \cref{lem:min-clique-cover}]
		Let $S$ be contained in temporal path $P$ in $\mathcal{T}$. Let $u_1, u_2, \ldots, u_{k}$ where $k=|S|$, be the ordering of the vertices in $S$ as they are encountered while traversing $P$ from the source to the sink. Notice that, for each $1\leq i < j\leq k$, there is a temporal path from $u_i$ to $u_j$. Therefore, $u_i$ is adjacent to $u_j$ in $G$. Hence, $S$ is contained in a clique of $G$.
		
		Let $S$ be contained in a clique of $G$ and $S'$ be a maximal complete subgraph of $G$ such that $S \subseteq V(S')$.
		Now, we orient the edges of $S'$ to  create a digraph $\overrightarrow{S'}$ as follows. 
		For an edge $uv\in E(S')$, we introduce an arc $\overrightarrow{uv}\in A(\overrightarrow{S'})$ if there is a temporal path from $u$ to $v$ in $\mathcal{T}$. Since $\mathcal{T}$ is acyclic, $\overrightarrow{S'}$ is a transitive tournament. 
		Hence, there is an ordering $u_1,u_2,\ldots,u_k$ of the vertices of $S'$ where $k=|V(S')|$ such that for $1\leq i<j\leq k$, there is a temporal path from $u_i$ to $u_j$ in $\mathcal{T}$. Now, consider any temporal path $P$ from $u_1$ to $u_k$ in $\mathcal{T}$. ($P$ exists as $\overrightarrow{u_1u_k}\in A(\overrightarrow{S'})$. Since $\mathcal{T}$ is a temporal oriented tree, $P$ will contain all vertices of $S'$ and therefore of $S$.
	\end{proof}
%\end{toappendix}

Following is an immediate corollary of the above.

\begin{corollary}\label{Cor-min-clique-cover}
	The minimum cardinality of a temporal path cover of $\mathcal{T}$ is equal to the minimum cardinality of a clique cover of $G$. 
\end{corollary}

We will often use the following lemma about vertex-intersecting temporal paths between pairs of vertices.
%\shortversion{We will often use the following lemma.}

\begin{lemma}%[*]
	\label{lem-intersectingPaths}
	Let $\{u,v,w,x\}\subseteq V(T)$ be four vertices such that any temporal path from $u$ to $v$ has a vertex in common with any temporal path from $w$ to $x$. 
	Then, there is a temporal path from $u$ to $x$, or a temporal path from $w$ to $v$.
\end{lemma}

%\begin{toappendix}
	\begin{proof}%[Proof of \cref{lem-intersectingPaths}]
		Assume that there is no temporal path from $u$ to $x$. 
		Let $y$ be the vertex of a temporal path from $w$ to $x$ that is closest to $u$ in $T$. Let $t$ be the smallest integer such that there is a temporal path from $u$ to $v$ that reaches $y$ at time-step $t$. Observe that no temporal path from $y$ to $x$ can start at time-step $t' > t$ since, otherwise, there would be a temporal path from $u$ to $x$.
		This implies that all temporal paths between $w$ and $x$ reach $y$ at time-step $t'' \leq t$.
		Let $P_1$ be a temporal path from $w$ to $y$ which is also a subpath of a temporal path from $w$ to $x$. Let $P_2$ be a temporal path from $y$ to $v$ which is also a subpath of a temporal path from $u$ to $v$. The above arguments imply that the arc incident with $y$ in $P_1$ has time-step at most $t$. Similarly, the arc incident with $y$ in $P_2$ has time-step strictly greater than $t$. Hence, the concatenation of $P_1$ and $P_2$ is a temporal path in $\mathcal{T}$ from $w$ to~$v$.
	\end{proof}
%\end{toappendix}

%\subsection{Forbidding holes and anti-holes}
\subsection{The case of holes}

%In this subsection, we shall show that the connectivity graph $G$ does not contain any holes or anti-holes. First, we prove the following lemma.
In this subsection, we will show that the connectivity graph $G$ does not contain any holes. We use the following lemma.

\begin{lemma}%[*]
	\label{obs:leaves}
	Let $H$ be an induced cycle of length at least $4$ in $G$. Then, for every vertex $v\in V(H)$ and every arc $\overrightarrow{a}$ of $T$ incident with $v$, the vertices of $H\setminus \{v\}$ lie in the same connected component of $T\setminus \{\overrightarrow{a}\}$.
\end{lemma}

%\begin{toappendix}
	\begin{proof}%[Proof of \cref{obs:leaves}]
		For the sake of contradiction, let there exist vertices $\{u,v,w\}\subseteq V(H)$ and an arc $\overrightarrow{a}$ of $T$ incident with $v$ such that $u$ and $w$ lie in two different connected components of $T'=T\setminus \{\overrightarrow{a}\}$. Let $C_u$ and $C_w$ be the sets of vertices of $H\setminus \{v\}$ contained in the same connected component as $u$ and  $w$, respectively. Since $H\setminus \{v\}$ is connected, there exist $u'\in C_u$ and $w'\in C_w$ such that $u'w'\in E(H)$ \emph{i.e.} $u'w'\in E(G)$. Hence, there is a temporal path $P$ from $u'$ to $w'$ or $w'$ to $u'$ in $\mathcal{T}$. Since $T$ is a tree, $P$ must contain $v$. \Cref{lem:min-clique-cover} implies that $\{u',v,w'\}$ forms a subset of a clique in $G$, and therefore $\{u',v,w'\}$ forms a triangle. But this contradicts that $H$ is a hole.
	\end{proof}
%\end{toappendix}

Going forward, we need the following notations. For an edge $e=uv\in E(G)$, let $Q_e$ denote a temporal path from $u$ to $v$ or $v$ to $u$ in $\mathcal{T}$. For an induced cycle $H$ of length at least $4$ in $G$, let $T_H$ denote the smallest connected subtree of $T$ containing all vertices of $H$. \Cref{obs:leaves} implies that every vertex of $H$ must be a leaf in $T_H$. For a vertex $v\in V(H)$, let $\overrightarrow{a}(v)$ be the arc incident with $v$ in $T_H$. 
Let $H$ be an induced cycle of length at least $4$ in $G$. %Based on the above notations,
We can partition the vertex set of $H$ into two sets $IN(H)$ and $OUT(H)$ as follows: a vertex $v\in V(H)$ is in $IN(H)$ if $\overrightarrow{a}(v)$ is directed towards $v$, and otherwise $v$ is in $OUT(H)$.

For a vertex $v\in IN(H)$, notice that both neighbors of $v$ in $H$ must lie in $OUT(H)$, and vice versa, since they must be connected by a directed path in $T$. 
Hence, $H$ is bipartite, and therefore $G$ does not contain any odd hole (\emph{i.e.}, a hole with an odd number of vertices):

\begin{lemma}\label{obs:no-odd-hole}
	The \connectivityGraph $G$ does not contain any odd hole.
\end{lemma}

Without loss of generality, we assume in the following that $OUT(H)$ (resp. $IN(H)$) contains every odd-indexed (resp. even-indexed) vertex of $H$.
For an even hole $H$ whose vertices are cyclically ordered as $u_1,u_2,\ldots,u_k$, we use a cyclic definition of, so $k+1=1$. 
We first prove the following lemmas.

\begin{lemma}%[*]
	\label{lem-consecutiveEdgesIntersect}
	Let $H$ be an even hole in the \connectivityGraph $G$. Then, for every $i$, $Q_{u_iu_{i+1}}$ and $Q_{u_{i+2}u_{i+3}}$ share a common vertex.
\end{lemma}

%\begin{toappendix}
	\begin{proof}%[Proof of \cref{lem-consecutiveEdgesIntersect}]
		Assume by contradiction that $Q_{u_iu_{i+1}}$ and $Q_{u_{i+2}u_{i+3}}$ are vertex-disjoint. Assume without loss of generality that $Q_{u_iu_{i+1}}$ goes from $u_i$ to $u_{i+1}$. 
		Note that, since each vertex of the hole is a leaf of $T_H$ as a consequence of \Cref{obs:leaves}, the two paths $Q_{u_iu_{i+1}}$ and $Q_{u_{i+1}u_{i+2}}$ have to share a common vertex other than $u_{i+1}$ (its neighbour in $T_H$). By the same reasoning, $Q_{u_{i+1}u_{i+2}}$ and $Q_{u_{i+2}u_{i+3}}$ share a common vertex other than $u_{i+2}$. Hence, since the three paths $Q_{u_iu_{i+1}}$, $Q_{u_{i+1}u_{i+2}}$ and $Q_{u_{i+2}u_{i+3}}$ are in $T_H$, and $Q_{u_iu_{i+1}}$ and $Q_{u_{i+2}u_{i+3}}$ are vertex-disjoint, there is an arc $\overrightarrow{a}$ contained in $Q_{u_{i+1}u_{i+2}}$ that separates $Q_{u_iu_{i+1}}$ and $Q_{u_{i+2}u_{i+3}}$.
		
		Removing $\overrightarrow{a}$ from $T$ partitions the vertices of $H$ into two sets $H_1$ and $H_2$: $H_1$ (resp. $H_2$) contains the vertices of $H$ that are in the same part of $T \setminus \overrightarrow{a}$ as $u_{i+1}$ (resp. $u_{i+2}$). Now, since $H$ is a cycle, there is an edge $u_ju_{j+1}$ such that (without loss of generality) $u_j \in H_1$, $u_{j+1} \in H_2$ and $(j,j+1) \neq (i+1,i+2)$. This implies that the path $Q_{u_ju_{j+1}}$ has to use $\overrightarrow{a}$ in $\mathcal{T}$, and thus $Q_{u_{i+1}u_{i+2}}$ and $Q_{u_ju_{j+1}}$ share a common vertex. Hence, \Cref{lem-intersectingPaths} implies that there is a temporal path from $u_{j+1}$ to $u_{i+1}$ or from $u_{i+2}$ to $u_j$. However, since $j \neq i+3$ ($u_j \in H_1$ and $u_{i+3} \in H_2$) and $j+1 \neq i$ ($u_{j+1} \in H_2$ and $u_i \in H_1$), both temporal paths would induce a chord in $H$, a contradiction.
	\end{proof}
%\end{toappendix}

\begin{lemma}%[*]
	\label{lem-noC6}
	The \connectivityGraph $G$ does not contain any hole of size 6.
\end{lemma}

%\begin{toappendix}
	\begin{proof}%[Proof of \Cref{lem-noC6}]
		Assume by contradiction that there is a hole on six vertices $u_1,\ldots,u_6$. We know that $Q_{u_1u_2}$ and $Q_{u_4u_5}$ are vertex-disjoint (since otherwise, by \Cref{lem-intersectingPaths}, at least one of the chords $u_1u_4$ or $u_2u_5$ would exist). The $u_i$'s are leaves of $T_H$, so $Q_{u_1u_2}$ and $Q_{u_1u_6}$, being paths with a common leaf in the same subtree, share at least one common vertex other than $u_1$ (its neighbour in $T_H$), let $v$ be the last (with respect to the orientation of $T$) vertex in their common subpath. Now, $Q_{u_5u_6}$ has a common vertex with both $Q_{u_1u_2}$ (by \Cref{lem-consecutiveEdgesIntersect}) and $Q_{u_1u_6}$ (the neighbour of $u_6$ in $T_H$), so it has to contain $v$ by the Helly property of subtrees of a tree. By the same reasoning, $Q_{u_4u_5}$ and $Q_{u_5u_6}$ share at least one common vertex other than $u_5$ (its neighbour in $T_H$), let $w$ be the last vertex in their common subpath. The Helly property of subtrees of a tree again implies that both $Q_{u_2u_3}$ and $Q_{u_3u_4}$ have to contain $w$, since they pairwise intersect with $Q_{u_4u_5}$. But this means that $Q_{u_2u_3}$ and $Q_{u_5u_6}$ share both $v$ and $w$ as common vertices, and so by \Cref{lem-intersectingPaths} there is at least one of the two chords $u_2u_5$ or $u_3u_6$, a contradiction.
	\end{proof}
%\end{toappendix}

We can now prove that there is no even hole in $G$:

\begin{lemma}
	\label{lem-noEvenHole}
	The \connectivityGraph $G$ does not contain any even hole.
\end{lemma}

\begin{proof}
	Assume by contradiction that $G$ contains an even hole $H$ on $k \geq 8$ vertices ($k=6$ is impossible by \Cref{lem-noC6}). We know by \Cref{lem-consecutiveEdgesIntersect} that both $Q_{u_3u_4}$ and $Q_{u_{k-1}u_k}$ intersect $Q_{u_1u_2}$, but do not intersect each other (otherwise, by \Cref{lem-intersectingPaths}, at least one of the edges $u_3u_k$ or $u_4u_{k-1}$ would exist, and both would be chords since $k \geq 8$), so there is an arc $\overrightarrow{a}$ in $T$ that separates them. Removing $\overrightarrow{a}$ from $T$ partitions the vertices of $H$ into two sets $H_1$ and $H_2$: $H_1$ (resp. $H_2$) contains the vertices of $H$ that are in the same part of $T \setminus \overrightarrow{a}$ as $u_3$ (resp. $u_{k}$). Now, since $H$ is a cycle, there is an edge $u_ju_{j+1}$ such that (without loss of generality) $u_j \in H_1$ and $u_{j+1} \in H_2$. This implies that the path $Q_{u_ju_{j+1}}$ has to use $\overrightarrow{a}$ in $\mathcal{T}$, and thus $Q_{u_1u_2}$ and $Q_{u_ju_{j+1}}$, both containing $\overrightarrow{a}$, share a common vertex. 
	Hence, \Cref{lem-intersectingPaths} implies that there is a temporal path from $u_{j+1}$ to $u_2$ or from $u_1$ to $u_j$. However, since $j \neq k$ ($u_j \in H_1$ and $u_k \in H_2$) and $j+1 \neq 3$ ($u_{j+1} \in H_2$ and $u_3 \in H_1$),	by \Cref{lem-intersectingPaths} both temporal paths would induce a chord in $H$, a contradiction.
\end{proof}

\subsection{The case of anti-holes}

\newcommand{\ODD}[1]{ODD\left(#1\right)}
\newcommand{\EVEN}[1]{EVEN\left(#1\right)}

In this subsection, we will show that the connectivity graph $G$ does not contain any anti-hole. 
For an anti-hole $H$, let its vertices be circularly ordered as $u_1,u_2,\ldots,u_k$ as they are encountered while traversing the complement of $H$ (which is a hole).
Let $\ODD{H}$ (resp. $\EVEN{H}$) denote the set of vertices with odd (resp. even) indices.

\begin{lemma}
	\label{lem-noAntihole}
	The \connectivityGraph $G$ does not contain any anti-hole.
\end{lemma}

\begin{proof}
	Assume by contradiction that $G$ contains an anti-hole $H$ with $k$ vertices. 
	If $k=5$, then $H$ is a hole, which contradicts \Cref{obs:no-odd-hole}; hence, assume $k\geq 6$.
	
	When $k$ is odd, let $F_1 = \ODD{H}\setminus \{u_k\}, F_2=\EVEN{H}$. When $k$ is even, let $F_1 = \ODD{H}, F_2=\EVEN{H}$. Observe that $|F_1|=|F_2|\geq 3$ and both sets induce (vertex-disjoint) cliques in $G$. By \Cref{lem:min-clique-cover}, there are temporal paths $P_1$ and $P_2$ in $\mathcal{T}$ containing $F_1$ and $F_2$, respectively, which we can assume are minimal vertex-inclusion-wise (so that, for each $i\in \{1,2\}$, both end-vertices of $P_i$ lie in $F_i$). For $i\in \{1,2\}$, let $v_i$ and $w_i$ denote the source and sink of $P_i$, respectively. 
	We have two cases.
	
	\medskip\noindent\textbf{Case 1: $V(P_1) \cap V(P_2) = \emptyset$. } Let $Q$ be the shortest temporal path that contains vertices from both $P_1$ and $P_2$. Let $p_1,p_2$ be the end-vertices of $Q$ that lie on $P_1$ and $P_2$, respectively. 
	Since for each $i\in \{1,2\}$ and $Z\in \{F_1,F_2\}$, $N_G(w_i)\cap Z\neq \emptyset$, $Q$ is oriented from $p_1$ to $p_2$, or vice versa. Without loss of generality, assume that $Q$ is oriented from $p_2$ to $p_1$. Then, necessarily $p_2= w_2$, since otherwise $w_2$ is not temporally connected with any vertex of $F_1$, a contradiction. 
	By a similar argument, we have $p_1=v_1$. 
	Now, consider the clique induced by $N(v_2)\cap F_1$. Due to \Cref{lem:min-clique-cover}, all vertices of $N(v_2)\cap F_1$ and $v_2$ itself are contained in a temporal path, which also necessarily contains $w_2$. Hence all of $F_2$ ($P_2$, even) is in a temporal path containing $v_1$, since the path has to go through $v_1$ to reach other vertices of $F_1$, and so $F_2\cup \{v_1\}$ forms a clique. This is a contradiction as $v_1$ necessarily has at least one non-neighbor in $F_2$.

	\medskip\noindent\textbf{Case 2: $V(P_1) \cap V(P_2) \neq \emptyset$. } Let $Q$ denote the maximal vertex-inclusion-wise path that is common to both $P_1$ and $P_2$, \emph{i.e.}, the path induced by the set $V(P_1) \cap V(P_2)$. Note that $Q$ does not contain any vertex from $H$, since a vertex of $H$ in $Q$ would be temporally connected to every other vertex of $F_1 \cup F_2$, a contradiction. Let $p$ denote source of $Q$ and for each $i\in \{1,2\}$ let $Q_i$ (resp. $Q'_i$) be the subpath of $P_i$ between $p$ and $w_i$ (resp. $p$ and $v_i$).
	
	Note that no vertex of $Q'_1\setminus\{p\}$ can be in a directed path with any vertex of $Q'_2\setminus\{p\}$. Similarly, no vertex of $Q_1\setminus\{p\}$ can be in a directed path with any vertex of $Q_2\setminus\{p\}$. Thus, the two subgraphs of the connectivity graph $G$ induced by the vertices of $(V(Q_1)\cup V(Q_2))\setminus V(Q)$ and $(V(Q'_1)\cup V(Q'_2))\setminus \{p\}$ each induce the complement of a complete bipartite graph.
	As $H$ does not contain any complement of a 4-cycle as an induced subgraph, this implies that there are exactly three vertices of $H$ in each of these two subsets of vertices (since $Q$ does not contain any vertex of $H$). In particular, $H$ has size either 6 or 7.
	
	Without loss of generality, we assume that $Q'_1$ contains only one vertex of $H$, which must be $v_1$. Thus, there are two vertices of $H$ in $Q'_2$: $v_2$ and another vertex, say, $v'_2$. Since $F_1$ and $F_2$ both have size~3, the vertices of $H$ in $Q_1$ are $w_1$ and (say) $w'_1$, and the only vertex of $H$ in $Q_2$ is $w_2$. Now, observe that if $v_2$ is contained in a temporal path with $w_1$, then $v_2$, $v'_2$, $w'_1$ and $w_1$ are in a common temporal path. This is not possible, since in $H$, there is either one or two non-edges among these four vertices (depending on whether $H$ has size 7 or 6). Thus, $w_1$ and $v_2$ are in no common temporal path. Since $v_2$ has no non-neighbour in $H$ other than $v_1$ and $w_1$, $v_2$ and $w'_1$ are in a common temporal path, that also contains $v'_2$. Thus, $\{v_2,v'_2,w'_1\}$ form a clique in $H$. Similarly, $\{v'_2,w'_1,w_1\}$ also form a clique in $H$. If $H$ had size~6, $v'_2$ and $w'_1$ would need to be non-neighbours in $H$ (since $w_1$ already has two non-neighbours in $H$), a contradiction. Thus, $H$ has size~7, and the two non-neighbours in $H$ of $u_7$ (the vertex of $H$ not in $F_1\cup F_2$) are $v'_2$ and $w'_1$ (since they are the only ones without two non-neighbours in $H$).
	But $u_7$ has to be temporally connected to all of $v_1$, $v_2$, $w_1$ and $w_2$, so $u_7$ has to be in $Q$. But any temporal path from $v_2$ to a vertex of $Q$ has to contain $v'_2$, and so $u_7$ and $v'_2$ are temporally connected, a contradiction. 
	This completes the proof.
\end{proof}

\subsection{Completion of the proof of \Cref{thm-orientedTrees}}
\Cref{obs:no-odd-hole,lem-noEvenHole,lem-noAntihole} imply that the connectivity graph of a temporal oriented tree is weakly chordal.
Note that this cannot be strengthened to chordal, as there are temporal oriented trees whose connectivity graphs contain induced $4$-cycles: let $\lambda(\overrightarrow{s_1c})=\lambda(\overrightarrow{s_2c})=1$ and $\lambda(\overrightarrow{ct_1})=\lambda(\overrightarrow{ct_2})=2$, the vertices $s_1$, $t_1$, $s_2$ and $t_2$ induce a $C_4$ in the \connectivityGraph.
\Cref{Cor-min-clique-cover} implies the correspondence between a minimum temporal path cover of a temporal oriented tree and a minimum clique cover of the corresponding \connectivityGraph. We then conclude using \Cref{thm:weakly} for the algorithm. \Cref{obs:max-ind-antichain}, \Cref{Cor-min-clique-cover} and \Cref{thm:weakly} together give the Dilworth property.

\section{\tempdcc on temporal oriented trees}
\label{sec-oriented}

\orientnp*

\begin{proof}
	The reduction is inspired from Theorem~1 in~\cite{temporalDP2,kunz2023graph}. However, in~\cite{temporalDP2,kunz2023graph}, the terminal vertices of the paths are fixed, which is not the case in our problem. Thus, nontrivial additions are needed. We reduce from \ubp, which is NP-complete~\cite{GareyJ79}. 
	
	\decisionpb{\ubp}{A list of item sizes $(x_1,\ldots,x_n)$, a number of bins $b$, each of size $B$. $x_1,\ldots,x_n,b,B$ are integers encoded in unary, and verify $\sum_{i=1}^{n} x_i = bB$.}{Is it possible to assign every item to a bin, filling all the bins?}{0.9}
	
	The idea of the reduction will be to have pairs of vertices serving as bins, each with $B$ leaves, and to have vertices representing, for each item, the bins that are unused by this item.
	
	We construct the following temporal oriented tree $\mathcal{T}=(T,\lambda)$:
	
	\begin{itemize}
		\item $V(T) = \{c\} \cup \mathlarger{\bigcup}_{i=1}^b \left\{ \{s_i,t_i\} \cup \mathlarger{\bigcup}_{j=1}^B \{r_i^j,u_i^j\} \right\} \cup \mathlarger{\bigcup}_{i=1}^n \mathlarger{\bigcup}_{j=1}^{(x_i-1)(b-1)} \{v_i^j,w_i^j\}$
		
		\item $A(T) = \mathlarger{\bigcup}_{i=1}^b \left\{ \{\overrightarrow{s_ic},\overrightarrow{ct_i}\} \cup \mathlarger{\bigcup}_{j=1}^B \{\overrightarrow{r_i^js_i},\overrightarrow{t_iu_i^j}\} \right\} \cup \mathlarger{\bigcup}_{i=1}^n \mathlarger{\bigcup}_{j=1}^{(x_i-1)(b-1)} \{\overrightarrow{v_i^jc},\overrightarrow{cw_i^j}\}$
	\end{itemize}
	
	For the sake of simplicity, we will use \emph{layers} to represent the time labels: for each item $i$, the layer $\lambda_i$ will assign to every arc a subset of $\{1,\ldots,2bx_i+4\}$. Thus, for an arc $a$, we have $\lambda(a)= \mathlarger{\bigcup}_{i=1}^n \{\ell + \sum_{j=1}^{i-1} (2bx_i+4)~|~\ell \in \lambda_i(a)\}$. This allows us to describe the layers starting with label~1. We call two time labels \emph{2-successive} if they differ by~2. The time labels in layer $i$ are as follows.
	
	\begin{itemize}
		\item For every $j \in \{1,\ldots,b\}$ and $k \in \{1,\ldots,B\}$: $\lambda_i(\overrightarrow{r_j^ks_j}) = \{\text{every 2-successive label between}\\2(j-1)x_i+1~\text{and}~2jx_i-1\}$; $\lambda_i(\overrightarrow{s_jc}) = \{\text{every 2-successive label between}~2(j-1)x_i+2~\text{and}~2jx_i\}$; $\lambda_i(\overrightarrow{ct_j}) = \{\text{every 2-successive label between}~2(j-1)x_i+3~\text{and}~2jx_i+1\}$; $\lambda_i(\overrightarrow{t_ju_j^k}) = \{\text{every}\\ \text{2-successive label between}~2(j-1)x_i+4~\text{and}~2jx_i+2\}$.
		\item For every $j \in \{1,\ldots,(x_i-1)(b-1)\}$: $\lambda_i(\overrightarrow{v_i^jc}) = \bigcup_{k=1}^b \{\text{the}~x_i-1~\text{first labels of}~\overrightarrow{ct_k}\}$; $\lambda_i(\overrightarrow{cw_i^j}) = \bigcup_{k=1}^b \{\text{the}~x_i-1~\text{highest labels of}~\overrightarrow{s_kc}\}$. For a given $k$, those are called the \emph{bin-$k$-labels}.
	\end{itemize}
	
	A layer of this construction is depicted in \Cref{fig-orientedTreesNP}. 
	The number of vertices is $1+2b+2bB+\sum_{i=1}^n (x_i-1)(b-1) = 2b(bB-n+1)+2n+1$, among which $2b(bB-n)+2n$ are leaves (half of them are sources, the other half sinks). The number of different time labels is $2bx_i+4$ for layer $i$, and thus a total of $2b(bB+4n)$. Hence, the reduction is polynomial.
	
	\begin{figure}
		\centering
		\scalebox{0.8}{\begin{tikzpicture}
				\node[noeud,scale=2] (c) at (5,4) {$c$};
				\node[noeud] (s1) at (0,7) {$s_1$};
				\node[noeud] (s2) at (2,7) {$s_2$};
				\node[noeud] (s3) at (4,7) {$s_3$};
				\node[noeud] (t1) at (0,1) {$t_1$};
				\node[noeud] (t2) at (2,1) {$t_2$};
				\node[noeud] (t3) at (4,1) {$t_3$};		
				\draw[\nicearrow] (s1)to node[midway,below,sloped]{$2,4,6$} (c);
				\draw[\nicearrow] (s2)to node[pos=0.4,below,sloped]{$8,10,12$} (c);
				\draw[\nicearrow] (s3)to node[pos=0.4,below,sloped]{$14,16,18$} (c);
				\draw[\nicearrow] (c)to node[midway,above,sloped]{$3,5,7$} (t1);
				\draw[\nicearrow] (c)to node[pos=0.6,above,sloped]{$9,11,13$} (t2);
				\draw[\nicearrow] (c)to node[pos=0.6,above,sloped]{$15,17,19$} (t3);
				\foreach \I in {1,2,3} {
					\foreach \J in {1,2,3,4} {
						\pgfmathsetmacro{\K}{0.25*\J+2*(\I-1)-0.625}
						\node[noeud] (r\I\J) at (\K,8) {};
						\node[noeud] (u\I\J) at (\K,0) {};
						\draw[\nicearrow] (r\I\J)to(s\I);
						\draw[\nicearrow] (t\I)to(u\I\J);
					}
					\pgfmathsetmacro{\L}{(\I-1)*2}
					\draw (\L,8.5) node {$r_\I^j$'s};
					\draw (\L,-0.5) node {$u_\I^j$'s};
				}
				\draw[rounded corners,dashed] (-0.625,7.75) rectangle (0.625,8.25);
				\draw[rounded corners,dashed] (1.375,7.75) rectangle (2.625,8.25);
				\draw[rounded corners,dashed] (3.375,7.75) rectangle (4.625,8.25);
				\draw[rounded corners,dashed] (-0.625,-0.25) rectangle (0.625,0.25);
				\draw[rounded corners,dashed] (1.375,-0.25) rectangle (2.625,0.25);
				\draw[rounded corners,dashed] (3.375,-0.25) rectangle (4.625,0.25);
				\draw (-0.75,7.5) node {1,3,5};
				\draw (-0.75,0.5) node {4,6,8};
				\draw (1.25,7.5) node {7,9,11};
				\draw (1.125,0.5) node {10,12,14};
				\draw (3.125,7.5) node {13,15,17};
				\draw (3.125,0.5) node {16,18,20};
				\foreach \I in {1,...,4} {
					\pgfmathsetmacro{\J}{6+0.4*(\I-1)}
					\node[noeud] (v1\I) at (\J,7) {};
					\node[noeud] (w1\I) at (\J,1) {};
					\draw[\nicearrow] (v1\I)to(c);
					\draw[\nicearrow] (c)to(w1\I);
				}
				\draw (7.5,5.5) node {3,5,9,11,15,17};
				\draw (7.5,2.5) node {4,6,10,12,16,18};
				\draw (6.6,7.5) node {$v_1^k$'s};
				\draw (6.6,0.5) node {$w_1^k$'s};
				\draw[rounded corners] (5.75,6.75) rectangle (7.45,7.25);
				\draw[dashed] (6.6,6.75)to(6.6,7.25);
				\draw[rounded corners] (5.75,0.75) rectangle (7.45,1.25);
				\draw[dashed] (6.6,0.75)to(6.6,1.25);
				\draw (8,7) node {$\mathbf{\ldots}$};
				\draw (8,1) node {$\mathbf{\ldots}$};
				\foreach \I in {1,...,4} {
					\pgfmathsetmacro{\J}{8.8+0.4*(\I-1)}
					\node[noeud] (vn\I) at (\J,7) {};
					\node[noeud] (wn\I) at (\J,1) {};
				}
				\draw (9.4,7.5) node {$v_n^k$'s};
				\draw (9.4,0.5) node {$w_n^k$'s};
				\draw[rounded corners] (8.55,6.75) rectangle (10.25,7.25);
				\draw[dashed] (9.4,6.75)to(9.4,7.25);
				\draw[rounded corners] (8.55,0.75) rectangle (10.25,1.25);
				\draw[dashed] (9.4,0.75)to(9.4,1.25);
		\end{tikzpicture}}
		\caption{Layer 1 of the reduction of the proof of \Cref{thm-orientedTreesNP}, with $x_1=3$, $b=3$, $B=4$. The only $v_j^k$'s and $w_j^k$'s linked with $c$ in this layer are those with $j=1$.}
		\label{fig-orientedTreesNP}
	\end{figure}
	
	We now prove that there is a valid assignment of every item to a bin if and only if there is a temporally disjoint path cover of $\mathcal{T}$ of size $b(bB-n)+n$.
	
	($\Rightarrow$) Suppose that $f: \{1,\ldots,n\} \rightarrow \{1,\ldots,b\}$ is a valid assignment (so $\sum_{i \in f^{-1}(j)} x_i = B$ for every $j$). In every layer $i$, we take $x_i$ \emph{$(r,u)$-paths} $(\overrightarrow{r_{f(i)}^js_{f(i)}},2(f(i)-1)x_i+2a), (\overrightarrow{s_{f(i)}c},\\2(f(i)-1)x_i+2a+1), (\overrightarrow{ct_{f(i)}},2(f(i)-1)x_i+2a+2), (\overrightarrow{t_{f(i)}u_{f(i)}^k},2(f(i)-1)x_i+2a+3)$ using uncovered $r_{f(i)}^j$'s and $u_{f(i)}^k$'s and for $a \in \{1,\ldots,x_i\}$; as well as all the paths from $v_i^j$ to $w_i^k$ such that the arc $\overrightarrow{v_i^jc}$ is used with a bin-$k$-label for $k \neq f(i)$. The first paths will always be possible, and will cover every $r_i^j$ and $u_i^j$ once all the layers are done, since we will use exactly $B$ paths for every $i$. The other paths can also clearly be constructed, since whenever $k \neq f(i)$ the bin-$k$-labels are not used by the first paths, and so we will cover all the $v_i^j$'s and $w_i^j$'s. Hence, we obtain a temporally disjoint path cover of $\mathcal{T}$ ($c$ and the $s_i$'s and $t_i$'s are clearly covered) of size $\sum_{i=1}^n (x_i + (x_i-1)(b-1)) = b(bB-n)+n$.
	
	($\Leftarrow$) Suppose that there is a temporal path cover of $\mathcal{T}$ of size $b(bB-n)+n$. The cover is of size twice the number of leaves, so each path in the cover will contain two leaves. Since every path between two leaves has to go through $c$ and the paths are temporally disjoint, there can be at most $bx_i$ paths in layer $i$, with equality if and only if all in-arcs with successive labels from $2$ to $2bx_i$ are used.
	
	The vertices $v_i^j$'s are linked with $c$ only in layer $i$, so they are covered in layer $i$ by an arc $\overrightarrow{v_i^jc}$ at time $d$ (note that $d$ is odd); sort them in a sequence $S$ ordered by the $d$'s. We call a subsequence of $S$ \emph{2-successive} if the $d$'s are 2-successive.
	
	%\begin{restatable}{claim}{OrientedTreesNPA}
	%	\label{clm-OrientedTreesNP1}
	%	Each 2-successive subsequence of length $k$ prevents $k+1$ $(r,u)$-paths from going through $c$.
	%\end{restatable}
	\begin{claim}%[*]
		\label{clm-OrientedTreesNP1}
		Each 2-successive subsequence of length $k$ prevents $k+1$ $(r,u)$-paths from going through $c$.
	\end{claim}
	
	%\shortversion{\noindent (Proof in the appendix due to space constraints.)}
	%\rtodo{replace with (*)}
%	\begin{toappendix}
		%\OrientedTreesNPA*
		
		\begin{proof}%[Proof of \Cref{clm-OrientedTreesNP1}]
			Each 2-successive subsequence $S'=(d,d+2,\ldots,d+2k-2)$ (recall that $d$ is odd) induces paths that occupy $c$ at all times between $d$ and $d+2k-1$ (since the last path needs to leave $c$ in order to allow another path to occupy $c$ in turn); but due to the difference of parity between the time labels of the arcs $\overrightarrow{u_i^jc}$ and $\overrightarrow{s_kc}$, $S'$ will prevent at least $|S'|+1=k+1$ $(r,u)$-paths from going through $c$, since no such path will be able to go through $c$ in the interval between $d-1$ and $d+2k-1$.
		\end{proof}
%	\end{toappendix}
	
	Note that having a path start in one layer and end in another layer will lower the maximum number of paths restricted to their layers by~1 for each of both, while gaining at most one path with arcs in two different layers.
	Furthermore, while the theoretical maximum number of paths in layer $i$ is $bx_i$, the $v_i^j$'s are covered in layer $i$, and thus by \Cref{clm-OrientedTreesNP1} there can be at most $bx_i - (b-1)$ paths in layer $i$. Since the number of paths in the cover is $b(bB-n)+n = \sum_{i=1}^n (bx_i - (b-1))$, there is no path having arcs in two different layers. Similarly, the only paths in the cover are $(r,u)$-paths and $(v,w)$-paths (since, otherwise, there would be even fewer paths in the layer).%\ftodo{between two layers $\to$ that contains arcs of two different layers?}\todo{A: Yes}
	
	%\begin{restatable}{claim}{OrientedTreesNPB}
	%	\label{clm-OrientedTreesNP2}
	%	There are exactly $x_i$ $(r,u)$-paths in layer $i$.
	%\end{restatable}
	\begin{claim}%[*]
		\label{clm-OrientedTreesNP2}
		There are exactly $x_i$ $(r,u)$-paths in layer $i$.
	\end{claim}
	
	%\shortversion{\noindent (Proof in the appendix due to space constraints.)}
%	\begin{toappendix}
		%\OrientedTreesNPB*
		
		\begin{proof}%[Proof of \Cref{clm-OrientedTreesNP2}]
			Due to the definition of $\lambda_i$, there are at least $b-1$ 2-successive subsequences, each of length at most $x_i-1$. By \Cref{clm-OrientedTreesNP1}, they prevent at least $(b-1)(x_i-1)+(b-1)=(b-1)x_i$ $(r,u)$-paths from going through $c$ in layer $i$ (since the number of $(v,w)$-paths is fixed, increasing the number of 2-successive subsequences will decrease their sizes, but will end up increasing the number of time labels during which $c$ is occupied). Since there are at most $bx_i$ paths per layer, there are at most $x_i$ $(r,u)$-paths in layer $i$, with equality if and only if there are exactly $b-1$ 2-successive subsequences. As this holds for every layer, the path cover is of size $b(bB-n)+n$, and $(b-1)(bB-n)$ paths are necessary to cover the $v_j^k$'s and $w_j^k$'s, there has to be $bB$ $(r,u)$-paths in all the layers; thus there are exactly $x_i$ such paths in layer $i$.
		\end{proof}
%	\end{toappendix}
	
	%\begin{restatable}{claim}{OrientedTreesNPC}
	%	\label{clm-OrientedTreesNP3}
	%	All the $(r,u)$-paths of a layer $i$ have to all go through the same $s_j$ and the same $t_{j'}$, with  $j=j'$.
	%\end{restatable}
	\begin{claim}%[*]
		\label{clm-OrientedTreesNP3}
		All the $(r,u)$-paths of a layer $i$ have to all go through the same $s_j$.
	\end{claim}
	
	%\shortversion{\noindent (Both proofs in the appendix due to space constraints.)}
	%\rtodo{replace with (*)}
	
%	\begin{toappendix}
		%\OrientedTreesNPC*
		\begin{proof}%[Proof of \Cref{clm-OrientedTreesNP3}]
			Assume by contradiction that there are $(r,u)$-paths in the same layer using $s_j$ and $s_k$ for $j \neq k$. Then, in order to cover the $v_i$'s and $w_i$'s, we need to have at least $b$ 2-successive subsequences (at least~1 for each of $j$ and $k$, and at least $b-2$ for the other bins). By \Cref{clm-OrientedTreesNP1} and as in the proof of \Cref{clm-OrientedTreesNP2}, this implies that there will be less than $x_i$ $(r,u)$-paths in the layer, which contradicts \Cref{clm-OrientedTreesNP2}, so all the $(r,u)$-paths go through the same $s_j$.
			
			The same argument can be applied to show that all $(r,u)$-paths go through the same $t_{j'}$, and that $j=j'$.
		\end{proof}
%	\end{toappendix}
	
	%Similarly, the $(r,u)$-paths of a layer will go through $s_j$ and $t_j$ with the same value of $j$.
	
	Hence, we can construct the item assignment function $f$ as follows: for every item $i$, let $j$ be the integer such that there are $x_i$ $(r,u)$-paths in layer $i$ going through $s_j$ and $t_j$; we define $f(i)=j$. \Cref{clm-OrientedTreesNP3} implies that $f$ will assign each item to exactly one bin. Moreover, by our construction, there are exactly $B$ $(r,u)$-paths going through each $s_j$, $x_i$ of which at layer $i$ for $i \in f^{-1}(j)$ by \Cref{clm-OrientedTreesNP2}, and thus it is a correct assignment: $\sum_{i \in f^{-1}(j)} x_i = B$ for each bin $j$.
\end{proof}

\longversion{We complement the above hardness result by the following construction.}
%\shortversion{We also show the following (proof in appendix due to space constraints).}
%\rtodo{replace with (*)}
We also show the following.

\begin{proposition}%[*]
	\label{prop-orientedStarTempDisjoint}
	There are \fprintorienteds (whose underlying digraph is a star) that do not satisfy the TD-Dilworth property.
\end{proposition}

%\begin{toappendix}
	%\star*
	\begin{proof}%[Proof of \Cref{prop-orientedStarTempDisjoint}]
		%\begin{proof}
		Consider the temporal oriented tree $\mathcal{S}_k=(S_k,\lambda)$, with $V(S_k)=\{s_1,\ldots,s_k\} \cup \{t_1,\ldots,t_k\} \cup \{c\}$, $A(S_k)=\bigcup_{i=1}^k \{\overrightarrow{s_iu},\overrightarrow{ut_i}\}$, and $\lambda(\overrightarrow{s_iu})=1$ and $\lambda(\overrightarrow{ct_i})=2$ for $i \in \{1,\ldots,k\}$. Now, as depicted on \Cref{fig-fprintoriented-partition-vs-antichain}, the $s_i$'s (or the $t_i$'s) form a (maximum-size) temporal antichain of size $k$. Since $c$ is a cut-vertex with all its in-arcs having the same time label, there can only be one path using $c$ to cover both an $s_i$ and a $t_i$, and thus every other vertex has to be covered individually. Hence, we need at least $2k-1$ temporal paths in any temporal path cover of $\mathcal{S}_k$ (and it is easy to construct such a cover). 
	\end{proof}
	
	\begin{figure}[h]
		\centering
		\begin{tikzpicture}
			\node[noeud] (s1) at (0,3) {};
			\node[noeud] (s2) at (1,3) {};
			\node[noeud] (sk) at (3,3) {};
			\node[noeud] (u) at (1.5,1.5) {};
			\node[noeud] (t1) at (0,0) {};
			\node[noeud] (t2) at (1,0) {};
			\node[noeud] (tk) at (3,0) {};
			
			\draw[line width = 0.25mm] \convexpath{s1,u,t1,u,s1}{0.24cm};
			\fill[white] \convexpath{s1,u,t1,u,s1}{0.23cm};
			\draw[line width = 0.25mm] (s2) circle (0.24cm);
			\draw[line width = 0.25mm] (sk) circle (0.24cm);
			\draw[line width = 0.25mm] (t2) circle (0.24cm);
			\draw[line width = 0.25mm] (tk) circle (0.24cm);
			
			\node[noeud] (s1) at (0,3) {};
			\node[noeud] (s2) at (1,3) {};
			\node[noeud] (sk) at (3,3) {};
			\node[noeud] (u) at (1.5,1.5) {};
			\node[noeud] (t1) at (0,0) {};
			\node[noeud] (t2) at (1,0) {};
			\node[noeud] (tk) at (3,0) {};
			
			\draw (2,3) node {$\mathbf{\ldots}$};
			\draw (2,0) node {$\mathbf{\ldots}$};
			
			\draw[\nicearrow] (s1)to node[midway,left,xshift=-2mm,yshift=-2mm]{1} (u);
			\draw[\nicearrow] (s2)to node[midway,right,yshift=1mm]{1} (u);
			\draw[\nicearrow] (sk)to node[midway,right,yshift=-1mm]{1} (u);
			\draw[\nicearrow] (u)to node[midway,left,xshift=-2mm,yshift=2mm]{2} (t1);
			\draw[\nicearrow] (u)to node[midway,right,yshift=-1mm]{2} (t2);
			\draw[\nicearrow] (u)to node[midway,right,yshift=1mm]{2} (tk);
			
			\draw (s1) node[above,yshift=2mm] {$s_1$};
			\draw (s2) node[above,yshift=2mm] {$s_2$};
			\draw (sk) node[above,yshift=2mm] {$s_k$};
			\draw (t1) node[below,yshift=-2mm] {$t_1$};
			\draw (t2) node[below,yshift=-2mm] {$t_2$};
			\draw (tk) node[below,yshift=-2mm] {$t_k$};
			\draw (u) node[right,xshift=2mm] {$c$};
		\end{tikzpicture}
		\caption{$\mathcal{S}_k$, a \fprintoriented with a maximum-size temporal antichain of size $k$ and a minimum-size \tempdccshort of size $2k-1$. }
		\label{fig-fprintoriented-partition-vs-antichain}
	\end{figure}
%\end{toappendix}

\section{Subclasses of \fprintorienteds}
\label{sec-rooted}

\longversion{In this section, we focus on subclasses of \fprintorienteds, showing that they satisfy the TD-Dilworth property and both \tempcc and \tempdcc can be solved efficiently for these classes.}

\rooted*

\longversion{
	\begin{algorithm}
		\caption{An algorithm for \tempcc on \fprintrooteds.}\label{alg-rootedTrees}
		\SetKwInOut{KwIn}{Input}
		\SetKwInOut{KwOut}{Output}
		\KwIn{A \fprintrooted $\mathcal{T}$.}
		\KwOut{A temporal path cover $\mathcal{C}$ of $\mathcal{T}$.}
		
		Sort the vertices of $\mathcal{T}$ with respect to their topological distance from the root in the underlying rooted directed tree (with the highest distances first)
		
		\While{there is an uncovered vertex}{
			Let $v$ be the first uncovered vertex
			
			Construct a maximum-length temporal path $P$ ending in $v$% and coming from vertices with lower labels
			
			Add $P$ to $\mathcal{C}$
		}
		
		Return $\mathcal{C}$
	\end{algorithm}
}

\begin{proof}
	$(a)$ Let $\mathcal{P}=(P,\lambda)$ be a temporal oriented line, and let $v$ be a leaf of $P$. We construct $\mathcal{C}$ as follows. Assume that $v$ is incident with an in-arc $\overrightarrow{uv}$. We construct a maximum-length temporal path that covers $v$. Set $(b,c)=(u,v)$, $\ell = \max \lambda(\overrightarrow{uv})$, and apply the following routine: while $b$ is incident with an in-arc $\overrightarrow{ab}$, if there is a time label smaller than $\ell$ in $\lambda(\overrightarrow{ab})$, add $\overrightarrow{ab}$ to the path, update $(b,c)=(a,b)$ and $\ell = \max \{k \in \lambda(\overrightarrow{ab})~|~k<\ell\}$. When the routine stops, add the path to $\mathcal{C}$, remove its vertices from $P$, and start again on a new leaf (or return $\mathcal{C}$ if $P$ is empty). If $v$ was incident with an out-arc, we would do the same but with out-arcs, start with the smallest possible time label, and update $\ell = \min \{k \in \lambda(\overrightarrow{ab})~|~k>\ell\}$.
	
	This algorithm computes its output in time $\mathcal{O}(\ell n)$: every arc is visited at most once, but we need to parse the time labels in order to see whether we can keep on extending the path or not.
	Furthermore, the set of leaves $v$ where we start the routine are a temporal antichain: assume on the contrary that $v_1$ and $v_2$ are such vertices that are temporally connected, and assume without loss of generality that there is a path from $v_1$ to $v_2$ in the underlying oriented path; in this case, our algorithm would have added $v_1$ to the path that started being computed at $v_2$, a contradiction. Hence, $\mathcal{C}$ is a temporally disjoint path cover with the same size as a temporal antichain, proving that it is minimum-size and that temporal oriented lines satisfy the TD-Dilworth property.
	
	$(b)$ We give an algorithm that solves \tempcc on a \fprintrooted $\mathcal{T}=(T,\lambda)$. First, we sort the vertices of $T$ with respect to their topological distance from the root in $T$ (with the highest distances first). Then, we construct a maximum-length temporal path covering the first uncovered vertex (which will be a sink of that path), until $\mathcal{T}$ is fully covered.
	
	Note that this algorithm outputs $\mathcal{C}$ which is clearly is a temporal path cover: every vertex is covered by some path of $\mathcal{C}$. Furthermore, it is an adaptation of the algorithm for temporal oriented lines: instead of successive leaves, we construct the paths from successive leaves with highest topological distance from the root. We will show that $\mathcal{C}$ is minimum-size, and later we will explain how to modify the algorithm in order to obtain a minimum-size \tempdccshort.
	
	Let $S$ be the set of sinks of paths of $\mathcal{C}$.
	First, let $v_i$ and $v_j$ be two vertices of $S$ (without loss of generality, assume that $v_i$ was covered by the algorithm after $v_j$). They cannot be temporally connected, since otherwise, the graph being a \fprintrooted, one of them is necessarily the predecessor of the other in a path from the root, and thus the maximum-length temporal path ending in $v_j$ would necessarily contain $v_i$, since there is a temporal path from $v_i$ to $v_j$, and thus $v_i$ would have been covered at this step and cannot be in $S$. Hence, $S$ is a temporal antichain.
	
	We now prove that $S$ is maximum-size. Assume by contradiction that there is a temporal antichain $S'$ with $|S'|>|S|$. However, by definition, no two vertices in $S'$ can be covered by the same temporal path of $\mathcal{C}$, and thus $|S'|=|\mathcal{C}|$. But $|\mathcal{C}|=|S|$, a contradiction.
	Thus, $S$ is a maximum-size temporal antichain of the \fprintrooted. Since the temporal antichain number is a lower bound for the temporal path cover number, this implies that the temporal path cover $\mathcal{C}$ that the algorithm constructed is minimum-size, and thus that \fprintrooteds satisfy the Dilworth property.
	
	We now modify the algorithm to obtain a minimum-size temporally disjoint path cover. Indeed, we can see that the maximum-length temporal path construction, which is executed for every vertex of $S$, can re-cover some vertices that had already been covered at a previous step. Let $v_i$ and $v_j$ be two vertices of $S$ such that their maximum-length temporal paths constructed by the algorithm $P_i$ and $P_j$ intersect.
	Since the graph is a \fprintrooted, we can divide $P_i$ and $P_j$ into the following parts, without loss of generality: $P_i = P_i^{\haut} \cup \left( P_i \cap P_j \right) \cup P_i^{\bas}$ and $P_j = \left( P_i \cap P_j \right) \cup P_j^{\bas}$, where $P_i^{\haut} \cap P_j = P_i^{\bas} \cap P_j = P_j^{\bas} \cap P_i = \emptyset$ (note that we can have $P_i^{\haut} = \emptyset$).
	Hence, we can modify the algorithm by adding a loop that, for each such pair $(P_i,P_j)$, defines those subpaths and then removes $P_i \cap P_j$ from $P_j$. Now, $\mathcal{C}$ will still be a temporal path cover, but the paths will be vertex-disjoint and thus temporally disjoint, and its size will not change. % Finally, note that, since $\mathcal{C}$ is vertex-disjoint, it is also temporally-disjoint. 
	This implies that \fprintrooteds satisfy the TD-Dilworth property (contrasting the general temporal oriented trees), and thus the modified algorithm outputs the optimal solution for those two problems.
	The result of the algorithm and its modification is depicted in \Cref{fig-fprintrooted-covering-and-partition}. \longversion{Note that none imposes any conditions on the labels.}
	
	Finally, one can check that the algorithm and its modification compute $\mathcal{C}$ in time $\mathcal{O}(\ell n^2)$. For each vertex in the antichain $S$, we have to construct the maximum-length temporal path. This can be done in time $\mathcal{O}(\ell n)$ by taking at every arc the largest label that allows to extend the path, thus we have to parse all the labels of every arc along the path, which can be of linear-size in the worst case. Since we can have a linear number of antichain vertices, we have a complexity of $\mathcal{O}(\ell n^2)$ to get the temporal path cover. The modification to make it temporally disjoint can be done in $\mathcal{O}(n^2)$ time afterwards.
\end{proof}

\begin{figure}[h]
	\centering
	\begin{subfigure}[b]{0.24\textwidth}
		\centering
		\scalebox{1}{
			\begin{tikzpicture}
				\node[noeud] (r) at (1,3) {};
				\node[noeud] (0) at (1,2) {};
				\node[noeud] (1) at (0,1) {};
				\node[noeud] (3) at (2,1) {};
				\node[noeud] (4) at (1.5,0) {};
				\node[noeud] (5) at (2.5,0) {};
				
				\draw[line width = 0.25mm] \convexpath{r,0,1,0,r}{0.24cm};
				\fill[white] \convexpath{r,0,1,0,r}{0.23cm};
				\draw[line width = 0.25mm] \convexpath{0,3,4,3,0}{0.22cm};
				\fill[white] \convexpath{0,3,4,3,0}{0.21cm};
				\draw[line width = 0.25mm] \convexpath{0,3,5,3,0}{0.15cm};
				\fill[white] \convexpath{0,3,5,3,0}{0.14cm};
				
				\node[noeud] (r) at (1,3) {};
				\node[noeud] (0) at (1,2) {};
				\node[noeud] (1) at (0,1) {};
				\node[noeud] (3) at (2,1) {};
				\node[noeud] (4) at (1.5,0) {};
				\node[noeud] (5) at (2.5,0) {};
				
				\draw[\nicearrow] (r)to node[midway,right,xshift=2mm] {2}(0);
				\draw[\nicearrow] (0)to node[midway,above left,xshift=-1mm,yshift=1mm] {3}(1);
				\draw[\nicearrow] (0)to node[midway,above right,xshift=1mm,yshift=1mm] {1}(3);
				\draw[\nicearrow] (3)to node[midway,left,xshift=-2mm,yshift=1mm] {2}(4);
				\draw[\nicearrow] (3)to node[midway,right,xshift=1mm,yshift=1mm] {2}(5);
			\end{tikzpicture}
		}
	\end{subfigure}
	\hfill
	\begin{subfigure}[b]{0.24\textwidth}
		\centering
		\scalebox{1}{
			\begin{tikzpicture}
				\node[noeud] (r) at (1,3) {};
				\node[noeud] (0) at (1,2) {};
				\node[noeud] (1) at (0,1) {};
				\node[noeud] (3) at (2,1) {};
				\node[noeud] (4) at (1.5,0) {};
				\node[noeud] (5) at (2.5,0) {};
				
				\draw[line width = 0.25mm] \convexpath{r,0,1,0,r}{0.24cm};
				\fill[white] \convexpath{r,0,1,0,r}{0.23cm};
				\draw[line width = 0.25mm] \convexpath{3,4,3}{0.24cm};
				\fill[white] \convexpath{3,4,3}{0.23cm};
				\draw[line width = 0.25mm] (5) circle (0.24cm);
				
				\node[noeud] (r) at (1,3) {};
				\node[noeud] (0) at (1,2) {};
				\node[noeud] (1) at (0,1) {};
				\node[noeud] (3) at (2,1) {};
				\node[noeud] (4) at (1.5,0) {};
				\node[noeud] (5) at (2.5,0) {};
				
				\draw[\nicearrow] (r)to node[midway,right,xshift=2mm] {2}(0);
				\draw[\nicearrow] (0)to node[midway,above left,xshift=-1mm,yshift=1mm] {3}(1);
				\draw[\nicearrow] (0)to node[midway,above right,xshift=-1mm,yshift=-1mm] {1}(3);
				\draw[\nicearrow] (3)to node[midway,left,xshift=-2mm,yshift=1mm] {2}(4);
				\draw[\nicearrow] (3)to node[midway,right,yshift=1mm] {2}(5);
			\end{tikzpicture}
		}
	\end{subfigure}
	\unskip\ \vrule\ 
	\begin{subfigure}{0.24\textwidth}
		\centering
		\scalebox{1}{
			\begin{tikzpicture}
				\node[noeud] (r) at (1,3) {};
				\node[noeud] (0) at (1,2) {};
				\node[noeud] (1) at (1,1) {};
				\node[noeud] (2) at (0,0) {};
				\node[noeud] (3) at (1,0) {};
				\node[noeud] (4) at (2,0) {};
				
				\draw[line width = 0.25mm] \convexpath{r,0,1,4,1,0,r}{0.24cm};
				\fill[white] \convexpath{r,0,1,4,1,0,r}{0.23cm};
				\draw[line width = 0.25mm] \convexpath{0,1,2,1,0}{0.19cm};
				\fill[white] \convexpath{0,1,2,1,0}{0.18cm};
				\draw[line width = 0.25mm] \convexpath{1,3,1}{0.16cm};
				\fill[white] \convexpath{1,3,1}{0.15cm};
				
				\node[noeud] (r) at (1,3) {};
				\node[noeud] (0) at (1,2) {};
				\node[noeud] (1) at (1,1) {};
				\node[noeud] (2) at (0,0) {};
				\node[noeud] (3) at (1,0) {};
				\node[noeud] (4) at (2,0) {};
				
				\draw[\nicearrow] (r)to node[midway,right,xshift=2mm] {1}(0);
				\draw[\nicearrow] (0)to node[midway,right,xshift=2.1mm] {1,2}(1);
				\draw[\nicearrow] (1)to node[midway,above left,xshift=-1mm,yshift=1mm] {2}(2);
				\draw[\nicearrow] (1)to node[midway,left,xshift=-1mm,yshift=-3mm] {1}(3);
				\draw[\nicearrow] (1)to node[midway,above right,xshift=1mm,yshift=1mm] {3}(4);
			\end{tikzpicture}
		}
	\end{subfigure}
	\hfill
	\begin{subfigure}{0.24\textwidth}
		\centering
		\scalebox{0.8}{
			\begin{tikzpicture}
				\node[noeud] (r) at (1,3) {};
				\node[noeud] (0) at (1,2) {};
				\node[noeud] (1) at (1,1) {};
				\node[noeud] (2) at (0,0) {};
				\node[noeud] (3) at (1,0) {};
				\node[noeud] (4) at (2,0) {};
				
				\draw[line width = 0.25mm] \convexpath{r,0,1,4,1,0,r}{0.24cm};
				\fill[white] \convexpath{r,0,1,4,1,0,r}{0.23cm};
				\draw[line width = 0.25mm] (2) circle (0.24cm);
				\draw[line width = 0.25mm] (3) circle (0.24cm);
				
				\node[noeud] (r) at (1,3) {};
				\node[noeud] (0) at (1,2) {};
				\node[noeud] (1) at (1,1) {};
				\node[noeud] (2) at (0,0) {};
				\node[noeud] (3) at (1,0) {};
				\node[noeud] (4) at (2,0) {};
				
				\draw[\nicearrow] (r)to node[midway,right,xshift=2mm] {1}(0);
				\draw[\nicearrow] (0)to node[midway,right,xshift=2mm] {1,2}(1);
				\draw[\nicearrow] (1)to node[midway,above left,xshift=1mm,yshift=-1mm] {2}(2);
				\draw[\nicearrow] (1)to node[midway,left,xshift=1mm] {1}(3);
				\draw[\nicearrow] (1)to node[midway,above right,xshift=1mm,yshift=1mm] {3}(4);
			\end{tikzpicture}
		}
	\end{subfigure}
	\caption{Minimum-size temporal path covers and temporally disjoint path covers of the same \fprintrooted (on the left, with one label per arc; on the right, with any labels per arc), as computed by our algorithm and its modification in the proof of \Cref{thm-rootedTrees}.}
	\label{fig-fprintrooted-covering-and-partition}
\end{figure}

\section{Algorithms for temporal digraphs of bounded treewidth}\label{sec-tw}

Recall that an algorithm is FPT with respect to some parameter $k$ of the input, if it runs in time $f(k)n^{\mathcal{O}(1)}$ for inputs of size $n$, where $f$ is any computable function; the algorithm is XP for this parameter if the running time is in $n^{f(k)}$~\cite{cygan2015}. We prove the following theorem. 

\FPTTempDisjoint*

%To prove the theorem, we use the concept of \emph{nice tree decompositions}~\cite{niceTW}. 
To prove the theorem, we use the well-known concept of \emph{nice tree decompositions}~\cite{niceTW}, which gives a very structured decomposition of a graph.% (full definition in the appendix). 

%\begin{toappendix}
	\begin{definition}
		A \emph{nice tree decomposition} of an undirected graph $G=(V,E)$ is a rooted tree $\mathtt{T}$ where each node $v$ is associated to a subset $X_v$ of $V$ called \emph{bag}, and each internal node has one or two children, with the following properties.
		\begin{enumerate}
			\item The set of nodes of $\mathtt{T}$ containing a given vertex of $G$ forms a nonempty connected subtree of $\mathtt{T}$.
			
			\item Any two adjacent vertices of $G$ appear in a common node of $\mathtt{T}$.
			
			\item Each node of $\mathtt{T}$ belongs to one of the following types: \emph{introduce}, \emph{forget}, \emph{join} or \emph{leaf}.
			
			\item A join node $v$ has two children $v_1$ and $v_2$ such that $X_v = X_{v_1} = X_{v_2}$.
			
			\item An introduce node $v$ has one child $v_1$ such that $X_v \setminus \set{x} = X_{v_1}$, where $x \in X_v$.
			
			\item A forget node $v$ has one son $v_{1}$ such that $X_v  = X_{v_1} \setminus \set{x}$, where $x \in X_{v_1}$.
			
			\item A leaf node $v$ is a leaf of $\mathtt{T}$ with $X_v=\emptyset$.
			
			\item The tree $\mathtt{T}$ is rooted at a leaf node $r$ with $X_r = \emptyset$.
		\end{enumerate}
	\end{definition}
%\end{toappendix}

It is known that for any undirected graph of treewidth $\tw$ with $n$ vertices, a tree-decomposition of width at most $2\tw$ can be computed in time $2^{\mathcal{O}(\tw)}n$~\cite{K21}, and the obtained tree decomposition can be transformed into a nice tree-decomposition of the same width with $\mathcal{O}(\tw n)$ bags in time $\mathcal{O}(\tw^2n)$~\cite{niceTW}.

\sloppy For the remainder of this section, we shall work with a temporal digraph $\mathcal D=(V,A_1,\ldots,A_{t_{\max}})$, %where every arc is active in at most $\ell$ time-steps, 
and a nice tree decomposition $\mathtt{T}$ of the underlying undirected graph of $D$. For a node $v$ of $\mathtt{T}$, let $\mathtt{T}_v$ denote the subtree of $\mathtt{T}$ rooted at $v$, and let $\mathcal D_v$ denote the temporal digraph induced by the union of the bags of nodes of $\mathtt{T}_v$. 
%We shall use the following observation in our proofs. 

The main idea behind the algorithm is to perform a bottom-up dynamic programming algorithm over $\mathtt{T}$. We can bound the number of partial solutions that can intersect a given bag, partly because of the following.

\begin{observation}\label{obs:arc_disjoint}
	Let $\mathcal{C}$ be a temporally disjoint path cover of $\mathcal D$. Then any arc of $\mathcal D$ appears in at most $\tmax$ many paths of $\mathcal{C}$.
\end{observation}

Consider an arbitrary temporally disjoint path cover $\mathcal{C}$ of $\mathcal D_v$. Observation~\ref{obs:arc_disjoint} implies that the number of temporally disjoint paths of $\mathcal{C}$ that contain at least one arc from the digraph induced by $X_v$ is at most the number of arcs in this digraph, times the number of time-steps at which each of the arcs is active in $\mathcal D$. This is at most $p={\tw\choose 2 } \cdot \tmax$. %Moreover, any vertex of $\mathcal D$ can be part of at most $t_{max}$ many temporal paths of $\mathcal{C}$.

Based on these, we create the following temporal multi-digraph. Let $\mathcal D'$ be a copy of $\mathcal D$. Now; for each arc $a$ with time labels $\lambda(a)=L\subseteq [\tmax]$, introduce $|L|$  many new arcs $a^1,\ldots,a^{|L|}$, each with a distinct time label of $L$. Note that any temporally disjoint path cover of $\mathcal D$ can be transformed into a temporally disjoint path cover of $\mathcal D'$ whose temporal paths are pairwise edge-disjoint. Therefore, from now on, we will consider $\mathcal D'$ instead of $\mathcal D$. 

We now describe the states of our dynamic programming algorithm. 
To do so, for a temporally disjoint path cover $\mathcal{C}$ of $\mathcal D$, its \emph{type} $\tau$ with respect to a node $v$ of $\mathtt{T}$ is determined by the following elements:

\begin{itemize}
	\item a partition $\mathcal Q=Q_0,Q_1,\ldots,Q_t$ of the arcs of $\mathcal D'$ inside $X_v$, where each part $Q_i$ corresponds to a temporal path $P(Q_i)$ of $\mathcal{C}$ (note that this path may form a set of disconnected sub-paths inside $X_v$), and where the part $Q_0$ is reserved for those arcs that do not belong to any path of $\mathcal C$;
	
	\item for each part $Q_i$ of $\mathcal Q$, the subset $V_i$ of vertices of $X_v$ that belong to $P(Q_i)$ (those are the endpoints of arcs in $Q_i$, together with those vertices of $P(Q_i)$ that are not incident with any arc in $Q_i$);
	
	\item for each part $Q_i$ of $\mathcal Q$, the order of the vertices of $V_i$ inside $P(Q_i)$, where $P(Q_i)$ is ordered from lowest to largest time label;
	
	\item for each part $Q_i$ of $\mathcal Q$, the set of vertices $x$ in $V_i$ with one or two arcs in $P(Q_i)$ from $x$ to a vertex $y$ not in $X_v$, together with the time labels of these arcs in $P(Q_i)$, and whether the arc connects $x$ to a vertex in $\mathcal D_v$ or in $\mathcal{D}\setminus \mathcal{D}_v$.
	
\end{itemize}

The total number of different types of solutions with respect to any node $v$ is at most $p^p \times 2^{\tw+1} \times (\tw+1)! \times 2^{\tw+2} \times \tmax^2$ which is $2^{\mathcal{O}(p\log p)}$. For a type $\tau$ with respect to node $v$, its \emph{size} is the bit-length of its encoding. A type $\tau$ with respect to $v$ is said to be \emph{consistent} if for each part $Q_i$ of $\mathcal Q$, there is a subset of vertices of $V_i$ whose ordering together with the arcs of $Q_i$, form a valid temporal path (in particular, all labels of $Q_i$ must be distinct). Moreover, the labels of required arcs from a vertex $x$ of $X_v$ to a vertex $y$ outside of $X_v$, must correspond to an actual arc label for some arc in $\mathcal D$ connecting $x$ to some vertex outside of $X_v$. Moreover, every vertex of $X_v$ must belong to some set $V_i$.
Whether a type $\tau$ with respect to $v$ is consistent can be checked in time proportional to $p$ and the size of $\tau$. For a node $v$ of $\mathtt{T}$ and a solution type $\tau$ with respect to $v$, let $opt(v,\tau)$ denote the minimum size of a solution for $\mathcal D_v$ that is of type $\tau$ with respect to $v$. The dynamic programming algorithm computes $opt(v,\tau)$ by traversing the nice tree-decomposition $\mathtt{T}$ bottom-up and computes, for each node $v$, all the values for $opt(v,\tau)$. The computation depends on whether the current node of $\mathtt{T}$ is a leaf, forget, introduce or join node. 

%\begin{toappendix}
	\medskip 
%	\paragraph*{Computation of $opt(v,\tau)$ on each node of the nice tree decomposition}
	
	\noindent \emph{Leaf node:} There is nothing to do since for a leaf node $v$, $X_v=\emptyset$, so there is no partial solution with respect to $v$.
	
	\medskip \noindent \emph{Forget node:} Let $v$ be a forget node that has a child node $v'$ such that $X_v=X_{v'}\setminus \{x\}$. For each possible consistent solution type $\tau$ with respect to $v$, we check which (consistent) solution types $\tau'$ with respect to $v'$ are compatible with $\tau$. Whether $\tau$ and $\tau'$ are compatible (meaning that, roughly speaking, $\tau$ corresponds to $\tau'$ by removing $x$) can be computed in time proportional to $p$, the size of $\tau$ and that of $\tau'$. Among those, we discard those where $x$ is required to have an arc to a vertex of $\mathcal{D}\setminus\mathcal{D}_v$ in its solution path (since $x$ will never appear again in the tree-decomposition). We let $opt(v,\tau)$ be the minimum value among all values $opt(v',\tau')$ with $\tau'$ one of the non-discarded types compatible with $\tau$.
	
	%Then the optimal solution for $\mathcal D_v$ is the minimum of all those solutions of $\mathcal D_{v'}$ where $w$ was not part of a singular path.
	
	\medskip \noindent \emph{Introduce node:} Let $v$ be an introduce node that has a child node $v'$ such that $X_v=X_{v'}\cup \{x\}$. For each possible consistent solution type $\tau$ with respect to $v$, we check which solution types with respect to $v'$ are compatible with it. Here, this means that $\tau$ can be obtained from $\tau'$ either as a new solution path with a single vertex, or by adding $x$ to one of the solution paths described by $\tau'$ (through an arc of the correct label as described in $\tau'$). If $x$ forms a single path in $\tau$, we let $opt(v,\tau)$ be the minimum over $opt(v',\tau')+1$, where $\tau'$ is compatible with $\tau$; otherwise, we take the minimum over all $opt(v',\tau')$ for compatible $\tau'$. 
	
	\medskip \noindent \emph{Join node:} Let $v$ be a join node with children $v_1$ and $v_2$ and $X_v=X_{v_1}=X_{v_2}$. For any three possible solution types $\tau$, $\tau_1$, $\tau_2$ that are consistent with respect to $v$, $v_1$ and $v_2$, respectively, we check if they are compatible. For this, the partitions of the arcs of $X_v$ have to agree, as well as the order of vertices inside each solution path. The other elements should also be compatible. For example, if in $\tau_1$ there is a vertex $x\in X_v$ that is required to have a single neighbour in $\mathcal D_{v_1}\setminus X_v$ in its solution path, and the same holds for $\tau_2$ and $\mathcal D_{v_2}\setminus X_v$, then the types are not compatible, since combining them would give two such neighbours to $x$ in $\mathcal D_{v}\setminus X_v$. Similarly, if in $\tau_1$, $x$ is required to have a neighbour in $\mathcal D_{v_1}\setminus X_v$ in its solution path and another neighbour in $\mathcal{D}\setminus\mathcal D_{v_1}$, $\tau_1$ is compatible with $\tau$ and $\tau_2$ if in $\tau_2$, $x$ is required to have a neighbour in $\mathcal D_{v_2}\setminus X_v$ in its solution path and another neighbour in $\mathcal{D}\setminus\mathcal D_{v_2}$, but in $\tau$, $x$ is required to have two neighbours in $\mathcal D_{v}\setminus X_v$ in its solution path. Other similar cases arise as well. For three compatible types $\tau$, $\tau_1$, $\tau_2$, $opt(v,\tau)$ is obtained from $opt(v_1,\tau_1)+opt(v_2,\tau_2)$ by subtracting the number of solution paths that intersect the bag (and that would otherwise be counted twice).
	%\medskip
%\end{toappendix}

This dynamic programming algorithm for \tempdcc takes $2^{\mathcal{O}(p\log p)}n$ time, which is an FPT running time of $2^{\mathcal{O}(\tw^2\tmax\log (\tw\tmax))}n$.

For \tempcc, the algorithm is similar, however, as the paths are not necessarily disjoint, the type of a solution with respect to $v$ must also contain the information of how many times a given part of $\mathcal{Q}$, representing a solution path with a certain intersection with the subgraph induced by $X_v$, appears in the solution $\mathcal C$. Thus, the number of possible solution types becomes $k^{\mathcal{O}(p\log p)}$, where $k$ is the solution size. We obtain a running time of $k^{\mathcal{O}(p\log p)}$, which is an XP running time of $n^{\mathcal{O}(\tw^2\tmax\log(\tw\tmax))}$ since we can assume $k\leq n$.

\section{Conclusion}\label{sec:conclu}

We have initiated the study of two natural path covering problems in temporal DAGs, which, in the static case, are related to Dilworth's theorem and are polynomial-time solvable. Both problems become NP-hard for temporal DAGs, even in a very restricted setting. Interestingly, and somewhat unexpectedly, they behave differently on temporal oriented trees: we showed that \tempcc is polynomial-time solvable on temporal oriented trees (and a temporal version of Dilworth's theorem holds in this setting), while \tempdcc remains NP-hard for this class.

{To prove our polynomial-time algorithm for \tempcc on temporal oriented trees, we have reduced the problem to \textsc{Clique Cover} in a static undirected graph, which turns out to be weakly chordal. This is a powerful technique, and the correspondence between the two problems is quite enlightening for the structure of temporal paths in an oriented tree. Nevertheless, it seems unlikely that this particular technique can be used on temporal digraph classes that are far from trees, as it was essential for the proof that any two vertices are joined by only one path in the underlying tree. However, this general technique could likely be applied in other temporal settings.}

We do not know if our algorithm for treewidth and number of time-steps is optimal. In particular, can we obtain an FPT algorithm for \tempcc for this parameter? One could also explore other (structural) parameterizations of the problems.

We note that many of our results for \tempdcc also hold for its stricter  vertex-disjoint version (note that a vertex-disjoint version of \textsc{Temporally Disjoint Paths} is studied in~\cite{KKK02-TPP}), in particular, the NP-hardness result for restricted DAGs and the polynomial-time algorithms for rooted directed trees and oriented lines.

\longversion{Recently, another version of disjoint paths in temporal graphs was introduced~\cite{ibiapina2023snapshot}. One could investigate the algorithmic behaviour of the corresponding path cover problem.}

\longversion{\ftodo{We can also ask about the largest possible ratio for TD-Dilworth property on trees: our example gives ratio 2, are there worse examples or not?}}

%\clearpage
\bibliographystyle{plainurl}
\bibliography{references}

%\newpage
%{\bf\Huge Appendix}
%\nopagebreak
\end{document}